\documentclass[onecolumn,superscriptaddress,nofootinbib,amsmath,amssymb,floatfix,float,aps,prd]{revtex4-2}
\usepackage{graphicx} %Include figure files
\usepackage{dcolumn} %Align table columns on decimal point
\usepackage{bm} %bold math
\usepackage[colorlinks=true, allcolors=blue]{hyperref}
\usepackage{cleveref}
\usepackage{xcolor}
\usepackage[normalem]{ulem}
\usepackage{babel}
\usepackage{booktabs}
\usepackage{threeparttable}
\usepackage{float}

\begin{document}

\title[]{Brans-Dicke-like field for co-varying $G$ and $c$: observational
constraints}

\author{J. Bezerra-Sobrinho}
\email{jeremias.bezerra.100@ufrn.edu.br}
\affiliation{Departamento de Física Teórica e Experimental, Universidade Federal do Rio
Grande do Norte, Campus Universitário--Lagoa Nova, Natal CEP 59078-970,
RN, Brazil}

\author{R. R. Cuzinatto}
\email{rodrigo.cuzinatto@unifal-mg.edu.br}
\affiliation{Department of Physics, University of Ottawa, Ottawa, ON K1N
6N5, Canada}
\affiliation{Instituto de Ciência e Tecnologia, Universidade Federal de
Alfenas, Rodovia José Aurélio Vilela 11999, Poços de Caldas CEP 37715-400,
MG, Brazil}

\author{L. G. Medeiros}
\email{leo.medeiros@ufrn.br}
\affiliation{Escola de Ciência e Tecnologia, Universidade Federal do Rio
Grande do Norte, Campus Universitário--Lagoa Nova, Natal CEP 59078-970,
RN, Brazil}

\author{P. J. Pompeia}
\email{pompeia@ita.br}
\affiliation{Departamento de Física, Instituto Tecnológico de Aeronáutica,
Praça Mal. Eduardo Gomes 50, São José dos Campos CEP 12228-900, SP,
Brazil}

\begin{abstract}
\vspace{0.2cm}
Ref. {[}\href{https://doi.org/10.3390/sym15030709}{Symmetry 15 (2023) 709}{]} introduced a Brans-Dicke-like framework
wherein the scalar field $\phi$ is composed of both $G$ and $c$
which, for this reason, co-vary according to $c^{3}/G=\text{constant}$.
In this paper, we use observational data to constrain the supposed
co-varying $G$ and $c$, under the hypothesis of the validity of the standard Lemaitre formula $1+z\sim a^{-1}$. The datasets include SN Ia, BAO and the value  of $\theta$ extracted from CMB data. A proxy
function is demanded for the varying $c$ since the framework does not
provide a closed set of equations for computing the functional form
of either $G$ or $c$ uniquely. Accordingly, we choose three
separate parameterizations for $c\left(z\right)$ inspired both by
desirable properties of the varying speed of light (VSL) and by successful phenomenological models from the literature---including the one by
Gupta (CCC framework in e.g. Ref. {[}\href{https://doi.org/10.1093/mnras/staa2472}{Mon. Not. R. Astron. Soc., 498 (2020) 4481-4491}{]}. When combined with DESI, Pantheon+ data strongly favor a variable speed of light with more than $3\sigma$ confidence level for all parameterizations considered in this paper, whereas Union2.1 suggests no variation of the speed of light. As we shall demonstrate, this apparent discrepancy is due to a strong correlation that emerges between $H_0$ and VSL.
\end{abstract}   

\maketitle

%\vspace{0.5cm}
%\tableofcontents{}
%\vspace{0.5cm}

\section{Introduction }\label{sec:Introduction}

General Relativity is our current best theory for describing gravity
\cite{Sabbata1986,Weinberg1972}. Its resulting cosmological model, the
$\Lambda$CDM model, is the benchmark model for our universe \cite{Weinberg1972,Ryden2017,Baumann2022}.
Still, there are some shortcomings---based both on theoretical features 
\cite{Bull2016,Weinberg1989}
 and on observational description limitations
\cite{Valentino2021,Handley2021,Vagnozzi2020,Ballardini2020,Braglia2021}---allowing some room for extensions. These
are broadly called modified gravity theories \cite{Capozziello2011,Petrov2020,Sotiriou2010,Faraoni2004}.

One possibility for such modification is the controversial proposal
of relaxing the constancy of physical couplings, such as the speed
of light $c$. Einstein himself admitted the latter possibility \cite{Einstein1907}.
Dirac was also bold in introducing his large-numbers hypothesis which
involved a cosmological scenario coming from a varying Newtonian coupling
$G$ \cite{Dirac1937,Dirac1938}. Brans and Dicke also admitted the
possibility of a varying effective $G$ through a scalar field $\phi$
whose Lagrangian should be added to the Einstein-Hilbert action \cite{BransDicke21961};
their goal was to fully realize Mach's principle, something that general
relativity falls short to implement \cite{Weinberg1972}. The idea of varying $c$ was revived in the early 1990's
by Barrow \cite{Barrow1998,Barrow1999}, and in the early 2000's by 
 Albrecht and Magueijo \cite{Albrecht1999}, with other notable contributors
in Refs. \cite{Ellis2005,Uzan2011}. The constraining of a supposed varying $c$ cosmology
was done in several works through the years \cite{Barrow1998,Barrow1999,Albrecht1999,Ellis2005,Uzan2011,Mendonca2021}.
Varying $G$ scenarios were too observationally restricted within several
contexts, e.g. those in Refs. \cite{Gupta2021,Holanda2025}. Some authors
even considered the possibility of a varying fine structure constant
$\alpha$, and fit their models to observational data to assess this
hypothesis on cosmological time scales \cite{Chakrabarti2022,Ferreira2024,Colaco2022,Colaco2019}.
Other researchers worked with models allowing the variation of two
or more couplings in the set $\left\{ G,c,\Lambda,\hbar,k_{B}\right\} $;
examples on this avenue include the works by Costa  et al. \cite{Costa2019}
and Nguyen \cite{Nguyen2025SNeIa,Nguyen2025h,Nguyen2025Higgs}, on the theoretical side; and
the papers by Lee \cite{Lee2021,Lee2023,Lee2023SNIa,Lee2024Co,Lee2024,Lee2025} and Gupta \cite{Gupta2020,Gupta2022,Gupta2022VCC,Gupta2023,Gupta2023Glob,Gupta2023JWST,Gupta2024,Gupta2024BAO},
more on the phenomenological vein---see also \cite{Cuzinatto2023CPC,Cuzinatto2022,Cuzinatto2023BHS} and references therein.

Brans-Dicke model features a dimensionless parameter $\omega$ in front
of the kinetic term for the scalar field $\phi=1/G$. This parameter
$\omega$ should be of order one for consistency. However, in the
results summarized by Will \cite{Will2018}, data constraining would demand $\omega\gtrsim4,000$. This
works against the Brans-Dicke theory. Then, one could argue that this proposal
should be discarded as unphysical. Still, Brans-Dicke theory is arguably
the paradigm of scalar-tensor theories in cosmology \cite{Faraoni2004}. Accordingly, other researchers would prefer to
modify or extend the Brans-Dicke theory to, perhaps, bypass the strict
constraints posed to the original model.

In the paper \cite{Cuzinatto2023}, we have proposed a Brans-Dicke-like
modified gravity proposal based on the action
\begin{align}
S & =\frac{1}{16\pi}\int d^{4}x\sqrt{-g}\left[\phi R-\frac{\omega}{\phi}\nabla_{\mu}\phi\nabla^{\mu}\phi-V\left(\phi\right)\right]\nonumber \\
 & +\int d^{4}x\sqrt{-g}\left[\frac{1}{c}\mathcal{L}_{m}\right],\label{eq:S_BD-like}
\end{align}
where $\omega$ is a dimensionless constant of order one. Eq. (\ref{eq:S_BD-like})
is formally the same as Brans-Dicke action \cite{Faraoni2004,BransDicke21961};
however, here the scalar field $\phi$ is defined as
\begin{equation}
\phi=\frac{c^{3}}{G};\label{eq:phi}
\end{equation}
it is a generalization of the original Brans-Dicke (BD) interpretation
in terms of the gravitational coupling $G$, namely $\phi_{\text{BD}}=1/G$.
The speed of light $c$ in Eq. (\ref{eq:phi}) is taken as a spacetime
function, just like $G$. The definition (\ref{eq:phi}) is based
on dimensionality consistency for $S$, which should be measured in
units of $\left(\text{energy}\right)\times\left(\text{time}\right)$;
the particular combination $c^{3}/G$ appears naturally also in the Einstein-Hilbert
action (where, of course, $c$ and $G$ are constants). We emphasize
that $c$ is a spacetime function in our approach and, for this reason,
the last term on the right-hand side of (\ref{eq:S_BD-like}) introduces
a coupling between the matter fields described by $\mathcal{L}_{m}$
and the varying speed of light. This is not the case in the standard Brans-Dicke
approach where the varying coupling is only $G$ and does not show
up in the matter term. The introduction of the factor $1/c$ in the last term of Eq.(\ref{eq:S_BD-like}) is not deduced from first principles, yet it is reasonable from dimensional analysis and accounting for the fact that $c$ is assumed as coordinate dependent. Moreover, this coupling between $c$ and $\mathcal{L}_{m}$ produces an interesting screening mechanism with phenomenological/observational consequences as discussed in \cite{Cuzinatto2025}.

By varying the action (\ref{eq:S_BD-like}) with respect to the (contravariant)
metric $g^{\mu\nu}$ and with respect to $\phi$, one gets the field
equations for the gravitational field and for the scalar field in
our Brans-Dicke-like approach---see Ref. \cite{Cuzinatto2023}. The
same reference performs the dynamical analysis of the model assuming
a physically reasonable functional form for the potential $V\left(\phi\right)$.
It is rigorously shown that the system evolves toward the equilibrium
point at which the scalar field in Eq. (\ref{eq:phi}) assumes a constant
value, 
\begin{equation}
\phi_{\text{eq}}=\frac{c_{0}^{3}}{G_{0}}=\text{constant}.\label{eq:phi_eq}
\end{equation}
At this point, the field equation for the gravitational field $g_{\mu\nu}$
resembles Einstein's equation of general relativity,\footnote{Once more, notice the coupling of the varying speed of light $c$
with the matter content described by $T_{\mu\nu}$ in the field equation
$\left(\ref{eq:FieldEq_gmunu}\right)$.} 
\begin{equation}
G_{\hphantom{\mu}\nu}^{\mu}=\frac{8\pi}{\phi_{\text{eq}}}\frac{1}{c}T_{\hphantom{\mu}\nu}^{\mu}-\Lambda_{0}\delta_{\nu}^{\mu}.\label{eq:FieldEq_gmunu}
\end{equation}
Here, the label 0 indicates the value of the parameter at present
cosmic time; $G_{\mu\nu}$ is the Einstein tensor of GR; $T_{\mu\nu}$
is the energy momentum tensor; $\Lambda$ is the cosmological constant
\cite{Weinberg1972}. Acting the covariant derivative onto (\ref{eq:FieldEq_gmunu}),
using the contracted Bianchi identity $\left(\nabla_{\mu}G_{\hphantom{\mu}\nu}^{\mu}=0\right)$,
and the metricity condition leads to the generalized conservation
law
\begin{equation}
\nabla_{\mu}\left(\frac{1}{c}T_{\hphantom{\mu}\nu}^{\mu}\right)=0,\label{eq:cov-conserv-eq}
\end{equation}
so that it is the effective energy momentum tensor $T_{\mu\nu}^{\left(\text{eff}\right)}=\frac{1}{c}T_{\mu\nu}$
that is actually covariantly conserved. 

The fact that $\phi=\phi_{\text{eq}}$ does not preclude the combined
variation of $G$ and $c$ but constrains them to follow $\frac{1}{\phi}\frac{d\phi}{dx^{0}}=-\left[\frac{1}{G}\frac{dG}{dx^{0}}-3\frac{1}{c}\frac{dc}{dx^{0}}\right]=0$ in a homogeneous and isotropic spacetime,
i.e. 
\begin{equation}
G=G_{0}\left(\frac{c}{c_{0}}\right)^{3}.\label{eq:G(c)}
\end{equation}
This means that if $c$ varies with respect to the ``time''-coordinate
$x^{0}=ct$, then it is mandatory that $G$ varies as well; and it
does so according to (\ref{eq:G(c)}). The latter was pointed out
by Gupta in \cite{Gupta2020} and explored in a number of papers
\cite{Gupta2021,Holanda2025,Cuzinatto2023,Cuzinatto2025}. In this paper, we will assess our Brans-Dicke-like model with covarying $G$ and $c$ from the observational data. It is important to highlight that this model does not allow us to determine the specific dynamics of $c$ (or $G$). For this, another degree of freedom should be introduced. In particular, a model in which $G$ and $c$ are treated as independent fields was explored in \cite{Cuzinatto2025}; there, the dynamics of both fields are constrained by a specific choice of the parameters of their kinetic terms. The present work can be interpreted as a reduction of this two-fields dynamical model caused by this constraint. Although restrictive, this hypothesis is used to bring about the phenomenological relation Eq. (\ref{eq:G(c)}) obtained by Gupta in other papers. In this context, the dependence of $c$ on $x^{0}$ is done by means of three different parameterizations that are constrained by observational data.

The remainder of the paper is organized as follows. In Section \ref{sec:Background-cosmology}
we specify the field equation (\ref{eq:FieldEq_gmunu}) for the homogeneous
and isotropic metric of background cosmology. We also discuss how
the varying speed of light (VSL) modifies the relation between the
temperature $T$ of the universe and its scale factor, showing that
the CMB spectrum is preserved. In order to make contact with observations,
we need check to how the cosmological distance measurements can be affected
by a varying speed of light; this is done in Section \ref{sec:Cosmo-Dist}---where
we build the expressions for proper distance, luminosity distance,
angular-diameter distance, sound horizon---and in Appendix \ref{app:Redshift}---where
it is shown that the redshift $z$ relation with the scale factor $a\left(t\right)$
of standard cosmology remains the same in our BD-like VSL model. This can be viewed as a simplifying hypothesis under the perspective that other works consider a modification of the standard Lemaitre formula in order to consider effects of the variation of the speed of light, namely the works by Nguyen \cite{Nguyen2025SNeIa,Nguyen2025Higgs,Nguyen2026}. Section
\ref{sec:Modelling-c(t)} presents the three parameterizations we
use for the varying $c$, putting then into context. Section \ref{sec:Observations}
finally constrains the three parameterizations for $c\left(z\right)$
using a combinations of various datasets, including SN Ia data (Pantheon$+$
and Union2.1) sets, Baryon Acoustic Oscillations (DESI), and CMB (Planck).
Our final remarks are given in Section \ref{sec:Final-remarks}.

\section{Background cosmology}\label{sec:Background-cosmology}

We will specify Eq. (\ref{eq:FieldEq_gmunu}) for the homogeneous
and isotropic line element of backgroung cosmology, the FLRW interval:
\begin{equation}
ds^{2}=-\left(dx^{0}\right)^{2}+a^{2}\left(x^{0}\right)\left[dr^{2}+S_{k}^{2}\left(r\right)\left(d\theta^{2}+\sin^{2}\theta d\phi^{2}\right)\right]\label{eq:FLRW(S_k)}
\end{equation}
where 
\begin{equation}
S_{k}\left(r\right)=\begin{cases}
R\sin\left(r/R\right)\,, & k=+1\\
r\,, & k=0\\
R\sinh\left(r/R\right)\,, & k=-1
\end{cases},\label{eq:S_k}
\end{equation}
with $R$ a distance parameter related to the curvature of the space
sector \cite{Ryden2017}. Notice that we carefully kept $x^{0}=ct$
for the sake of covariance, since now $c$ is in principle a function
of the cosmic time $t$. It means that 
\begin{equation}
dx^{0}=c\left(t\right)dt.\label{eq:dx0(c,dt)}
\end{equation}

Plugging the metric components from the line element (\ref{eq:FLRW(S_k)})
into the field equations (\ref{eq:FieldEq_gmunu}) alongside $T_{\hphantom{\mu}\nu}^{\mu}=\text{diag}\left\{ -\varepsilon,p,p,p\right\} $
($\varepsilon$ and $p$ are the energy density and pressure of the
perfect fluid respectively) leads to the Friedmann equations (in terms
of $x^{0}$): 
\begin{equation}
\bar{H}^{2}=\frac{8\pi}{3\phi_{\text{eq}}}\frac{1}{c}\varepsilon+\frac{\Lambda_{0}}{3}-\frac{k}{a^{2}},\label{eq:FriedEq(x0)}
\end{equation}
and 
\begin{equation}
\frac{d\bar{H}}{dx^{0}}+\bar{H}^{2}=-\frac{4\pi}{3\phi_{\text{eq}}}\frac{1}{c}\left(\varepsilon+3p\right)+\frac{\Lambda_{0}}{3}.\label{eq:AccelEq(x0)}
\end{equation}
Here, we define the Hubble function as 
\begin{equation}
\bar{H}=\frac{1}{a}\frac{da}{dx^{0}},\label{eq:H-bar(x0)}
\end{equation}
i.e. the derivative of the scale factor $a$ is with respect to $x^{0}$
(not w.r.t. the cosmic time $t$).\footnote{Notice the change in notation for the Hubble function in this
paper with respect to that in Ref. \cite{Cuzinatto2023}. Here, we
use $\bar{H}$ while Ref. \cite{Cuzinatto2023} uses simply $H$.
In that reference, the cosmic time $t$ is never used so there is no
reason for confusion; herein, we will reintroduce $t$ and the distinction
with the standard Hubble function $H=\frac{1}{a}\frac{da}{dt}$ is
necessary.}

The (generalized) continuity equation stems from the Noether theorem
applied to the energy momentum tensor of a perfect fluid \cite{Cuzinatto2023},
cf. Eq. (\ref{eq:cov-conserv-eq}): 
\begin{equation}
\frac{d\varepsilon}{dx^{0}}+3\bar{H}\left(\varepsilon+p\right)=\frac{1}{c}\frac{dc}{dx^{0}}\varepsilon.\label{eq:ContEq(x0)}
\end{equation}
For a constant $c=c_{0}$, the right-hand side of this equation vanishes
and we recover the energy conservation of standard cosmology (with
$x^{0}=c_{0}t$).

In what follows, we will seek contact with observations. For this reason,
it is necessary to express the basic equations of background cosmology
in terms of the cosmic time $t$ rather than in terms of the $x^{0}$
coordinate. From the point of view of the FLRW line element (and of
the metric components), a varying $c$ could be seen as a time reparameterization
due to Eq. (\ref{eq:dx0(c,dt)}). However, the speed of light does
not show up only as a space-time causality coupling in the time sector
of the line element, $dx^{0}=c_{\text{ST}}dt$ (the label ST stands
for ``spacetime''). In fact, the speed of light is present in multiples
areas of Physics, including electromagnetism (EM), where $c_{\text{EM}}=\left(\varepsilon_{0}\mu_{0}\right)^{-1/2}=c_{0}$
($\varepsilon_{0}$ is the electric permittivity of vacuum; $\mu_{0}$
stands for the vacuum magnetic permeability). We do not want to tamper
with laboratory results of terrestrial scale experiments involving
$c_{\text{EM}}$; at the same time, we wish to allow an evolution
of $c_{\text{ST}}=c\left(t\right)$ over cosmological time spans in
order to study the possible impacts of our co-varying $c$ and $G$
for cosmic evolution. This will motivate modeling $c=c\left(t\right)$
as a function starting from $c_{0}$ in the early universe, exhibiting
a nontrivial behavior $c=c\left(t\right)$ until recently in the cosmic
history, and returning to the value $c_{0}=3\times10^{8}\text{ m/s}$
around the present-day time $t_{0}$. We will talk more about this
in Section \ref{sec:Modelling-c(t)}; here, this hypothesis is mentioned
to allay concerns about possible dramatic modifications of Big-Bang
Nucleosynthesis (BBN) and CMB features (in the early universe) or
local experiments involving electromagnetism (in the present universe).

According to the previous paragraph, it is instrumental to rewrite 
Eq. (\ref{eq:FriedEq(x0)}) using (\ref{eq:dx0(c,dt)}): 
\begin{equation}
\bar{H}=\frac{1}{a}\frac{da}{dt}\frac{dt}{dx^{0}}=\frac{1}{c\left(t\right)}H,\label{eq:H-bar(H)}
\end{equation}
where $a=a\left(x^{0}\left(t\right)\right)$ and 
\begin{equation}
H=\frac{1}{a}\frac{da}{dt}.\label{eq:H}
\end{equation}
is the usual Hubble function. Therefore, due to (\ref{eq:phi_eq}),
Eq. (\ref{eq:FriedEq(x0)}) reads: 
\begin{equation}
H^{2}=H_{0}^{2}\left[\frac{8\pi G_{0}}{3c_{0}^{2}H_{0}^{2}}\left(\frac{c}{c_{0}}\right)\left(\varepsilon_{d}+\varepsilon_{b}+\varepsilon_{\gamma}+\varepsilon_{\nu}\right)+\frac{\Lambda_{0}c_{0}^{2}}{3H_{0}^{2}}\left(\frac{c}{c_{0}}\right)^{2}\right].\label{eq:H(epsilon_i,L)}
\end{equation}
Herein, the labels $d$, $b$, $\gamma$ and $\nu$ refer to dark matter, baryons, photons and neutrinos, respectively
The standard definitions 
\begin{equation}
E\equiv\frac{H}{H_{0}}\label{eq:E(H)}
\end{equation}
and 
\begin{equation}
\Omega_{\Lambda}\equiv\frac{\Lambda_{0}c_{0}^{2}}{3H_{0}^{2}},\qquad\Omega_{i}\equiv\frac{\varepsilon_{i}}{\varepsilon_{c}},\qquad\text{with}\qquad\varepsilon_{c}\equiv\frac{3H_{0}^{2}c_{0}^{2}}{8\pi G_{0}}\label{eq:Omega_i}
\end{equation}
are particularly useful here. Those cast Eq. (\ref{eq:H(epsilon_i,L)})
into the form 
\begin{equation}
E\left(a\right)=\sqrt{\left[\frac{c\left(a\right)}{c_{0}}\right]\left[\Omega_{m}\left(a\right)+\Omega_{r}\left(a\right)\right]+\left[\frac{c\left(a\right)}{c_{0}}\right]^{2}\Omega_{\Lambda}},\label{eq:E(Omega_i(a))}
\end{equation}
where $\Omega_{m}=\Omega_{\text{d}}+\Omega_{b}$ and $\Omega_{r}=\Omega_{\gamma}+\Omega_{\nu}$
are the matter and the radiation-like particles density parameters.

Because $a=a\left(x^{0}\right)$, the continuity equation, Eq. (\ref{eq:ContEq(x0)}),
is the same as 
\begin{equation}
\frac{d\varepsilon_{i}}{da}+\frac{3}{a}\left(\varepsilon_{i}+p_{i}\right)=\frac{1}{c}\frac{dc}{da}\varepsilon_{i}\label{eq:ContEq(epsilon,p)}
\end{equation}
where we have used (\ref{eq:H-bar(x0)}) and assumed that each one
of the $i$-th background component is conserved separetely $\left(i=\left\{ m,r,\Lambda\right\} \right)$.
The pressure is described by an equation of state (EoS) of the type
\begin{equation}
p_{i}=w_{i}\varepsilon_{i},\label{eq:EoS}
\end{equation}
with $w_{m}=0$, $w_{r}=1/3$, and $w_{\Lambda}=-1$; in any case
$w_{i}=\text{constant}$. Inserting (\ref{eq:EoS}) into (\ref{eq:ContEq(epsilon,p)}),
leads to: 
\begin{equation}
\frac{d}{da}\left(\frac{\varepsilon_{i}}{c}\right)+\frac{3}{a}\left(\frac{\varepsilon_{i}}{c}\right)\left(1+w_{i}\right)=0.\label{eq:ContEq(w)}
\end{equation}
This is immediately integrated to: 
\begin{equation}
\frac{\varepsilon_{i}}{c}=\left(\frac{\varepsilon_{i0}}{c_{0}}\right)a^{-3\left(1+w_{i}\right)},\label{eq:epsilon_i(a,c)}
\end{equation}
with $a_{0}=a\left(t_{0}\right)=1$ and $\varepsilon_{i0}=\varepsilon_{i}\left(a_{0}\right)$.
Specifically, 
\begin{equation}
\frac{\varepsilon_{m}}{c}=\frac{\varepsilon_{m0}}{c_{0}}a^{-3};\qquad\frac{\varepsilon_{r}}{c}=\frac{\varepsilon_{r0}}{c_{0}}a^{-4};\qquad\frac{\varepsilon_{\Lambda}}{c}=\text{constant}.\label{eq:epsilon_m,r,L}
\end{equation}
Interestingly, the evolution of the density depends on the varying
$c$ in our BD-like VSL model. The matter energy density $\varepsilon_{m}$
not only scales with $a^{-3}$ (as in standard cosmology) but also
increases with $c$. Analogously, $\varepsilon_{r}$ and $\varepsilon_{\Lambda}$
exhibit their expected behaviors from standard cosmology times a
factor $c$. 

Eq. (\ref{eq:epsilon_m,r,L}) can be substituted into (\ref{eq:E(Omega_i(a))})
{[}if we recall the definition $\Omega_{i}=\varepsilon_{i}/\varepsilon_{c}$
in (\ref{eq:Omega_i}){]}. The result is: 
\begin{equation}
E\left(a\right)=\frac{c\left(a\right)}{c_{0}}\sqrt{\Omega_{m0}a^{-3}+\Omega_{r0}a^{-4}+\Omega_{\Lambda}}.\label{eq:E(a)}
\end{equation}
For purposes of data fitting, it is more convenient to express Eq.
(\ref{eq:E(a)}) in terms of the redshift. Appendix \ref{app:Redshift}
shows that, under our hypothesis, the expression redshift $z$ as a function of the scale
factor $a$ is \emph{not} altered by a varying speed of light, i.e.
\begin{equation}
\left(1+z\right)=\frac{1}{a}\label{eq:z(a)}
\end{equation}
holds even in our Brans-Dicke-inspired co-varying $c$ (and $G$)
framework. The demonstration of (\ref{eq:z(a)}) in Appendix \ref{app:Redshift}
is very similar to the standard derivation of the cosmological redshift (see e.g. Refs. \cite{Weinberg1972,Ryden2017}); however, we felt it was a necessary task in view of the fact that other VSL frameworks admit a dependence of the redshift with respect to both $a$ and $c$---one
example is the meVSL framework by S. Lee \cite{Lee2025,Lee2021},
and references therein. A second example is the proposal by Nguyen \cite{Nguyen2025SNeIa,Nguyen2026}. In these works, the value of $c$ inside the host galaxy of an emitting SN Ia is not necessarily the same $c$ inside the Milky Way, both of which are different from $c$ in the expanding cosmic void. This modifies the standard Lemaitre formula by introducing explicitly all these different regimes of the varying $c$ (the last part of Appendix \ref{app:Redshift} presents details on this more general approach) From (\ref{eq:z(a)}) into (\ref{eq:E(a)}):
\begin{equation}
E\left(z\right)=\frac{c\left(z\right)}{c_{0}}\sqrt{\Omega_{m0}\left(1+z\right)^{3}+\Omega_{r0}\left(1+z\right)^{4}+\Omega_{\Lambda}}.\label{eq:E(z)}
\end{equation}

We derived the photon energy density as a function of both the scale
factor and the varying speed of light in Eq. (\ref{eq:epsilon_m,r,L});
it reads $\varepsilon_{\gamma}\sim c\left(a\right)a^{-4}$. This is
different from the usual result $\varepsilon_{\gamma}\sim a^{-4}$.
Not surprisingly, this difference leads to the violation of the standard
relation $T\sim a^{-1}$ relating the scale factor and the temperature
of gas of photons in the universe. In fact, statistical mechanics
teaches us that 
\begin{equation}
\varepsilon_{\gamma}=\frac{\pi^{2}k_{B}^{4}}{15\hbar^{3}c^{3}}T^{4}.\label{eq:StefBoltz}
\end{equation}
This is the Stefan-Boltzmann law \cite{Pathria2011}. It will be assumed
valid not only in present-day times but also throughout cosmic history.
We emphasize that $\hbar=h/2\pi$ ($h$ is Planck's constant) and $k_{B}$ are regarded constant in this paper.\footnote{A discussion regarding the possible variation of $k_{B}$ can be found
on page 22 of Ref. \cite{Cuzinatto2023}.}

Inserting Eq. (\ref{eq:StefBoltz}) into Eq. (\ref{eq:ContEq(w)})
with $w_{r}=1/3$, gives: 
\begin{equation}
\frac{d}{da}\left(\frac{T}{c}\right)^{4}=-\frac{4}{a}\left(\frac{T}{c}\right)^{4}\label{eq:DiffEqT(a)}
\end{equation}
The above differential equation is solved by 
\begin{equation}
T\left(a\right)=T_{0}\frac{c}{c_{0}}\frac{a_{0}}{a}.\label{eq:T(a)}
\end{equation}
This recovers $T\sim a^{-1}$ for $c=\text{constant}$. However, a
varying $c$ modifies the way we perceive the evolution of cosmic
temperature. As a consequence, the redshift of photon decoupling $z_{\text{dec}}$
changes with respect to the prediction of standard cosmology; the
redshift of electron freeze-out $z_{d}$ will also change in our framework.
We could prevent this by requiring that $c$ changes only in between
after CMB emission and before recent times. In this way, we guarantee
that the thermodynamics at recombination remains unaltered: $T_{\text{dec}}\sim0.26\text{ eV}\sim3000\text{ K}$
and $T_{d}\sim0.25\text{ eV}\sim2885\text{ K}$.

It is also paramount to assure that CMB observations still hold, leading
to $T_{0}\sim2.361\times10^{-4}\text{ eV}\sim2.725\text{ K}$. It
is a concern if that would actually be possible in our context since
Eq. (\ref{eq:T(a)}) violates the ordinary relation $T\sim1+z$ for
$c=c\left(a\right)$. It can be show, however, that there is no reason
for worries in this regard. In effect, the cosmic background radiation
generated on the occasion of the last scattering surface exhibits a
blackbody spectrum of the form\footnote{\textcolor{black}{There is a coefficient depending on (a power) of $\nu$ in the formula for the black-body spectrum. It does not matter
for the discussion of the spectrum preservation in VSL models: being
a multiplicative factor, it is able to modify the overall amplitude
of the spectrum, but not its general shape.}} 
\begin{equation}
R\propto\frac{1}{e^{h\nu/kT}-1}.\label{eq:L_blackbody}
\end{equation}
where $\nu=c/\lambda$ and $R$ is the radiance. The relation between
the radiation wavelength $\lambda$ and the scale factor $a$ is still
given by 
\begin{equation}
\frac{\lambda}{a}=\frac{\lambda_{0}}{a_{0}}\label{eq:lambda(a)}
\end{equation}
in the context of our VSL framework {[}cf. Eq. (\ref{eq:l_0(l_e)})
in Appendix \ref{app:Redshift}.{]} Now, Eqs. (\ref{eq:T(a)}) and
(\ref{eq:lambda(a)}) lead to: 
\begin{equation}
\frac{hc}{\lambda kT}=\frac{ah}{\lambda k}\frac{c}{aT}=\frac{a_{0}h}{\lambda_{0}k}\frac{c_{0}}{a_{0}T_{0}}=\frac{hc_{0}}{\lambda_{0}kT_{0}}.\label{eq:blackbody-factor}
\end{equation}
This is the factor appearing in the blackbody spectrum (\ref{eq:L_blackbody}),
so that the radiance 
\[
R\propto\frac{1}{e^{hc/\lambda kT}-1}=\frac{1}{e^{hc_{0}/\lambda kT_{0}}-1}\propto R_{0}
\]
remains unchanged by the varying $c$ in an expanding universe. This
is true only because the temperature obeys the new relation (\ref{eq:T(a)}).\footnote{Notice that the $c$ appearing in the argument of the exponential
of the blackbody spectrum is the speed of light from electromagnetism
$c=c_{\text{EM}}$. Our treatment also makes use of $c_{\text{ST}}$,
the speed of light in the line element $ds^{2}$, which is interpreted
as the causality coupling. For us, $c=c_{\text{EM}}=c_{\text{ST}}$.}

\section{Contact with observations: cosmological distances }\label{sec:Cosmo-Dist}

In this section we develop the observables connecting our varying speed
of light scenario with cosmological observations.

\subsection{Proper distance}\label{subsec:Proper-distance}

From the observational viewpoint, it is key to determine the proper
distance $d_{p}$. Most of the derivation process for obtaining the
expression for $d_{p}$ can be adapted from the steps clearly described
in Ref. \cite{Ryden2017}---see also \cite{Gupta2020}.

A fixed-time spatial geodesic for constant $\theta$ and $\phi$ is
described by the interval (\ref{eq:FLRW(S_k)}) with $dt=d\theta=d\phi=0$:
$ds=a\left(t\right)dr$. The \emph{proper distance} $d_{p}$ between
the observer (at $r=0$) and the source (at radial position $r$)
is: 
\begin{equation}
d_{p}\left(t\right)=\int_{\text{observer}}^{\text{\text{source}}}ds=a\left(t\right)\int_{0}^{r}dr^{\prime}=a\left(t\right)r\label{eq:d_p(r)}
\end{equation}
{[}Eq. (20) in \cite{Gupta2020}.{]} The proper distance today is:
\begin{equation}
d_{p}\left(t_{0}\right)=a\left(t_{0}\right)r\label{eq:d_p(t_0)}
\end{equation}

Now, the $r$ is the distance covered by a photon (for which $ds=0$)
since the emission at time $t_{e}$ to the detection at $t_{0}$.
In a radial trajectory, $\theta$ and $\phi$ are constant, so that
(\ref{eq:FLRW(S_k)}) leads to: 
\begin{equation}
0=-c^{2}\left(t\right)dt^{2}+a^{2}\left(t\right)dr^{2}\Rightarrow\int_{0}^{r}dr^{\prime}=\int_{t_{e}}^{t_{0}}\frac{c\left(t\right)}{a\left(t\right)}dt\Rightarrow r=\int_{t_{e}}^{t_{0}}\frac{c\left(t\right)}{a\left(t\right)}dt\label{eq:r(t)}
\end{equation}
Plugging (\ref{eq:r(t)}) into (\ref{eq:d_p(t_0)}): 
\begin{equation}
d_{p}\left(t_{0}\right)=a\left(t_{0}\right)\int_{t_{e}}^{t_{0}}\frac{c\left(t\right)}{a\left(t\right)}dt\,.\label{eq:d_p(c,a)}
\end{equation}
This equation correponds to Eq. (21) in Ref. \cite{Gupta2020}.

Eqs. (\ref{eq:d_p(c,a)}) can be cast in terms of the redshift $z$.
The scale factor $a=a\left(t\right)$ is assumed to be a monotonically
increasing function of time. Hence, there exists the inverse function
$t=t\left(a\right)$ and 
\begin{equation}
\frac{dt}{dz}=\frac{dt}{da}\frac{da}{dz}=\frac{1}{aH}\left(-\frac{a^{2}}{a_{0}}\right)=-\frac{\left(\frac{a}{a_{0}}\right)}{H_{0}\left(\frac{H}{H_{0}}\right)}=-\frac{1}{H_{0}}\frac{1}{\left(1+z\right)}\frac{1}{E\left(z\right)}\label{eq:dtdz}
\end{equation}
Inserting (\ref{eq:z(a)}) and (\ref{eq:dtdz}) into (\ref{eq:d_p(c,a)}):
\[
d_{p}\left(t_{0}\right)=a\left(t_{0}\right)\int_{z_{e}}^{0}\frac{c\left(t\right)}{a\left(t_{0}\right)}\left(1+z\right)\left[-\frac{1}{H_{0}}\frac{1}{\left(1+z\right)}\frac{1}{E\left(z\right)}dz\right]
\]
where $t=t_{0}\rightarrow z=0$ and $t=t_{e}\rightarrow z=z_{e}$.
(The quantity $z_{e}$ is the redshift at the photon's emission.)
Also, $c\left(t\right)=c\left(t\left(z\right)\right)$. Hence, 
\begin{equation}
d_{p}\left(t_{0}\right)=\frac{1}{H_{0}}\int_{0}^{z}\frac{c\left(z^{\prime}\right)}{E\left(z^{\prime}\right)}dz^{\prime}.\label{eq:d_p(z,c)}
\end{equation}
The explicit dependence of the integral kernel on the varying $c$
appears to modify the expression for the proper distance with respect
to that in standard cosmology. However, if one substitutes Eq. (\ref{eq:E(z)})
into (\ref{eq:d_p(z,c)}), 
\begin{equation}
d_{p}\left(t_{0}\right)=\frac{c_{0}}{H_{0}}\int_{0}^{z}\frac{dz^{\prime}}{\sqrt{\Omega_{m0}\left(1+z^{\prime}\right)^{3}+\Omega_{r0}\left(1+z^{\prime}\right)^{4}+\Omega_{\Lambda}}},\label{eq:d_p(t_0,z)}
\end{equation}
one notices that $c\left(z\right)$ vanishes, thus concluding that
(\ref{eq:d_p(t_0,z)}) is the same as in the conventional case, i.e.
$d_{p}=\left.d_{p}\right|_{\left\{ G,c\right\} =\text{constant}}$.

Eq. (\ref{eq:d_p(z,c)}) matches Eq. (20) of Ref. \cite{Cuzinatto2022}.
However, in the latter case, the Co-varying Physical Couplings (CPC)
framework produces a modified version of Eq. (\ref{eq:d_p(t_0,z)}),
because its Friedmann equation (providing the function $E=H/H_{0}$)
is not the same as Eq. (\ref{eq:H(epsilon_i,L)}) of our BD-like co-varying
$G$ and $c$ scenario. Contrary to what happens here, the proper
distance in the CPC framework is not as in standard cosmology. This
makes it clear that different scenarios involving variable $G$ and
$c$ will demand separate analysis of the proper distance formula
and other observable quantities.

\subsection{Luminosity distance}\label{subsec:Luminosity-distance}

Luminosity $L$ is defined as energy $E$ per unit time $\delta t$:
$L=E/\delta t$. The energy spreads out radially from the source in
spherical wave fronts of area $A=4\pi d_{L}^{2}$; here, the radial
distance $d_{L}$ is defined as the \emph{luminosity distance}. The
flux $f$ is a measure of the distribution of the luminosity over
$A$: 
\begin{equation}
f=\frac{L}{4\pi d_{L}^{2}}\label{eq:f(d_L)}
\end{equation}
The above definition is inspired by Euclidean geometry in a static
universe. When extrapolated to a dynamical universe, it is usual to
understand that $f=f_{0}$ is computed at the present time ($t=t_{0}$,
hence the label $_{0}$) and $L=L_{e}$ is the luminosity given off
by the source (label $_{e}$: emitted).

The quantity $d_{L}$ will be determined by the comparison of (\ref{eq:f(d_L)})
with the analogous expression taking into account both the expansion
of the universe (through $a\left(t\right)$ or $z$) and the varying
speed of light $c\left(t\right)$. For the FLRW geometry, we have
$A=4\pi S_{k}^{2}\left(r\right)$ since $S_{k}\left(r\right)$ is
the comoving measure of distance---see Eq. (\ref{eq:FLRW(S_k)});
specifically, a flat space section with $S_{k=0}\left(r\right)=r$
gives $A=4\pi r^{2}$; cf. Eq. (\ref{eq:S_k}).

The luminosity $L$ also changes in an expanding universe due to the
modification of the photon energy. Planck's formula $E=h\nu$ ($\nu$ is the frequency of the photon of
wavelength $\lambda$ propagating with speed $c=\lambda\nu$) leads
to: 
\begin{equation}
E_{0}=\frac{hc_{0}}{\lambda_{0}}=\frac{\lambda_{e}}{\lambda_{0}}\frac{hc_{e}}{\lambda_{e}}\frac{c_{0}}{c_{e}}=\frac{\lambda_{e}}{\lambda_{0}}\frac{c_{0}}{c_{e}}E_{e}\Rightarrow E_{0}=\frac{1}{\left(1+z\right)}\frac{c_{0}}{c_{e}}E_{e}.\label{eq:E_0(z,c,E_e)}
\end{equation}
In the last step we used the definition of redshift $z\equiv\left(\lambda_{0}-\lambda_{e}\right)/\lambda_{e}$.
In this regard, check also Eqs. (\ref{eq:l_0(l_e)}) and (\ref{eq:z(a_e)})
in Appendix \ref{app:Redshift}. Incidentally, Eq. (\ref{eq:l_0(l_e)})
relates the wavelength and the scale factor; it allows us to conclude
that 
\begin{equation}
\frac{\lambda_{e}}{a_{e}}=\frac{\lambda_{0}}{a_{0}}\Rightarrow\frac{c_{e}\delta t_{e}}{a_{e}}=\frac{c_{0}\delta t_{0}}{a_{0}}\Rightarrow\delta t_{0}=\left(1+z\right)\frac{c_{e}}{c_{0}}\delta t_{e};\label{eq:dt_0(z,c)}
\end{equation}
meaning that the time interval $\delta t$ between two wave crests
at detection is modified with respect to the same quantity at emission
due to (i) the expansion of the universe and (ii) the varying $c$.
Due to Eqs. (\ref{eq:E_0(z,c,E_e)}) and (\ref{eq:dt_0(z,c)}) the
flux in FLRW universe is computed as: 
\[
f_{0}=\frac{L_{0}}{4\pi S_{k}^{2}\left(r\right)}=\frac{\left(E_{0}/\delta t_{0}\right)}{4\pi S_{k}^{2}\left(r\right)}=\frac{1}{\left(1+z\right)^{2}}\left(\frac{c_{0}}{c_{e}}\right)^{2}\frac{\left(E_{e}/\delta t_{e}\right)}{4\pi S_{k}^{2}\left(r\right)},
\]
where $L_{e}=E_{e}/\delta t_{e}$. Therefore, 
\begin{equation}
f_{0}=\frac{L_{e}}{4\pi\left[S_{k}\left(r\right)\left(1+z\right)\frac{c\left(z\right)}{c_{0}}\right]^{2}},\label{eq:f_0(z,c)}
\end{equation}
where $c_{e}=c\left(z\right)$. By confronting Eqs. (\ref{eq:f(d_L)})
and (\ref{eq:f_0(z,c)}), we obtain the luminosity density in our
BD-like co-varying $G$ and $c$ scheme: $d_{L}\left(z\right)=S_{k}\left(r\right)\left(1+z\right)\frac{c\left(z\right)}{c_{0}}$.
As stated previously, $S_{k}\left(r\right)=r$ for the flat space geometry
of $k=0$. Moreover, the radial distance $r$ from the observer to
the source is estimated as the proper distance at the present-day
time, $d_{p}\left(t_{0}\right)$.\footnote{In fact, from Eq. (\ref{eq:d_p(t_0)}): $d_{p}\left(t_{0}\right)=a\left(t_{0}\right)r$.
With the normalization $a\left(t_{0}\right)=a_{0}=1$, we have $r=d_{p}\left(t_{0}\right)$.} Hence, $S_{k}\left(r\right)=d_{p}\left(t_{0}\right)$ for all practical
purposes in this paper, and we write finally: 
\begin{equation}
d_{L}\left(z\right)=d_{p}\left(t_{0}\right)\left(1+z\right)\frac{c\left(z\right)}{c_{0}}.\label{eq:d_L(z,c)-flat}
\end{equation}
Notice that $d_{p}\left(t_{0}\right)$ is also a function of $z$,
since it is given by (\ref{eq:d_p(t_0,z)}). Eq. (\ref{eq:d_L(z,c)-flat})
shows that the luminosity distance in our varying-$c$ scenario is
different from $d_{L}$ as predicted by standard cosmology. Model
constraining via SNe Ia data depends on $d_{L}$; therefore, we expect
that this dataset will be the key in telling our BD-like cosmology apart
from $\Lambda$CDM cosmology.

\subsection{Angular-diameter distance}\label{subsec:Angular-diameter-distance}

The \emph{angular-diameter distance} $d_{A}$ of an object of standardized
length $\ell$ can be computed via the small-angle formula $\ell=d_{A}\delta\theta$
of Euclidean geometry \cite{Cuzinatto2023BHS}. The small angle $\delta\theta$
is perceived by the observer at a comoving distance $r$ from the
object at one single instant of time. For this reason, $dr=dt=0$;
moreover, we are talking about a length (not a patch) in the sky,
so that $d\phi=0$. In an expanding FLRW universe, $\ell$ is the
proper distance between the two end-points of the standard yardstick:
$\ell=ds$ with $\delta\theta=d\theta$. Therefore, from Eq. (\ref{eq:FLRW(S_k)}),
\begin{equation}
\ell=a\left(t_{e}\right)S_{k}\left(r\right)d\theta=\frac{S_{k}\left(r\right)}{\left(1+z\right)}d\theta\label{eq:proper-length}
\end{equation}
with $t=t_{e}$ corresponding to the instant of light emission by
the standard yardstick and $\left(1+z\right)=1/a_{e}$, as usual---Eq.
(\ref{eq:z(a_e)}). Comparison with $\ell=d_{A}\delta\theta$, gives:
\begin{equation}
d_{A}=\frac{S_{k}\left(r\right)}{\left(1+z\right)}.\label{eq:d_A(S_k,z)}
\end{equation}
This reprodces the result in standard cosmology \cite{Ryden2017}.
This fact was expected: By taking $dt=0$ in Eq. (\ref{eq:FLRW(S_k)}),
we eliminate the $c=c\left(t\right)$-dependence of $\ell$.

For the flat space geometry, $S_{k=0}\left(r\right)=d_{p}\left(t_{0}\right)$.
Then Eq. (\ref{eq:d_A(S_k,z)}) is written as: 
\begin{equation}
d_{A}=\frac{d_{p}\left(t_{0}\right)}{\left(1+z\right)},\label{eq:d_A(z,c)-flat}
\end{equation}
with $d_{p}\left(t_{0}\right)$ given by Eq. (\ref{eq:d_p(t_0,z)}).

\subsection{Distances related to the sound horizon: CMB and BAO}\label{subsec:Sound-horizon-BAO}

The goal of this section is to build the equations with which the
baryon acoustic oscillations (BAO) are analyzed in the context of
a varying $c$ (and $G$) model. We are also interested in the angular
acoustic scale $\theta_{\ast}$ obtained from the position of the
first peak of the CMB power spectrum. These quantities are built from
the sound horizon \cite{Baumann2022}, 
\begin{equation}
r_{s}\left(t\right)=\int_{0}^{t}\frac{c_{s}\left(t^{\prime}\right)}{a\left(t^{\prime}\right)}dt^{\prime}\Rightarrow r_{s}\left(z\right)=\int_{z}^{\infty}\frac{c_{s}\left(z^{\prime}\right)}{H\left(z^{\prime}\right)}dz^{\prime}=\frac{c_{0}}{H_{0}}\int_{z}^{\infty}\frac{c_{s}\left(z^{\prime}\right)}{c_{0}}\frac{dz^{\prime}}{E\left(z^{\prime}\right)},\label{eq:sond-horizon}
\end{equation}
computed at the appropriate time (or their corresponding redshifts).
Eq. (\ref{eq:sond-horizon}) contains the speed of sound $c_{s}$,
the rate of propagation of acoustic waves in the medium. For example,
the last scattering surface is related to 
\begin{equation}
r_{*}=r_{s}\left(z_{*}\right)=\frac{c_{0}}{H_{0}}\int_{z_{*}}^{\infty}\frac{c_{s}\left(z\right)}{c_{0}}\frac{dz}{E\left(z\right)},\label{eq:r_star}
\end{equation}
which is the sound horizon at photon decoupling, $z_{\ast}\simeq1090$.
The angular acoustic scale is a function of both $r_{*}$ and the
proper distance $d_{p}$ at the redshift of photon decoupling \cite{Baumann2022}:
\begin{equation}
\theta_{\ast}=\frac{r_{\ast}}{d_{A}\left(z_{\ast}\right)},\label{eq:theta_star}
\end{equation}
where is $d_{A}\left(z_{\ast}\right)$ given by Eq. (\ref{eq:d_A(z,c)-flat})
at $z=z_{*}$:
\begin{equation}
d_{A}\left(z_{*}\right)=\frac{c_{0}}{H_{0}(1+z_{*})}\int_{0}^{z_{*}}\frac{c\left(z\right)}{c_{0}}\frac{dz}{E\left(z\right)}.\label{eq:d_p(z_star)}
\end{equation}

Before the epoch of last scattering, baryons and photons are strongly
coupled through Compton scattering. In this situation, the baryonic
fluid experiences a competition between the gravitational pull and
radiation pressure. This competition produces acoustic waves of matter
propagating at speed $c_{s}\left(t\right)$. The maximum comoving
distance covered by these matter waves defines the sound horizon at
the drag epoch, $r_{d}=r_{s}\left(z_{d}\right)$, which is computed
by 
\begin{equation}
r_{d}=r_{s}\left(z_{d}\right)=\frac{c_{0}}{H_{0}}\int_{z_{d}}^{\infty}\frac{c_{s}\left(z\right)}{c_{0}}\frac{dz}{E\left(z\right)},\label{eq:r_d(z_d)}
\end{equation}
where $z_{d}\simeq1060$ is the redshift of the baryon drag epoch,
or simply the $z$-drag.

The speed of sound $c_{s}\left(z\right)$ is \cite{Baumann2022}:
\begin{equation}
c_{s}\left(z\right)=\frac{c\left(z\right)}{\sqrt{3\left[1+\frac{3}{4}\frac{\Omega_{b}\left(z\right)}{\Omega_{\gamma}\left(z\right)}\right]}}.\label{eq:c_s(z)}
\end{equation}
In standard cosmology (of non-varying $G$ and $c$), the speed of
light is constant and the numerator of (\ref{eq:c_s(z)}) exhibits
$c_{0}$ instead of $c\left(z\right)$. In varying speed of light
scenarios, however, the speed of light depends explicitly on the redshift:
$c=c\left(z\right)$. For this reason, one would naïvely expects that
the sound horizon (\ref{eq:sond-horizon}) would also be $c\left(z\right)$-dependent.
As it turns out, however, this is not the case in our model. In order
to see this, consider first the ratio $\left(\Omega_{b}/\Omega_{\gamma}\right)$
in the denominator of (\ref{eq:c_s(z)}): 
\begin{equation}
\frac{\Omega_{b}\left(z\right)}{\Omega_{\gamma}\left(z\right)}=\frac{\varepsilon_{b}}{\varepsilon_{\gamma}}=\frac{\frac{c}{c_{0}}\frac{\varepsilon_{m0}}{\varepsilon_{c}}a^{-3}}{\frac{c}{c_{0}}\frac{\varepsilon_{r0}}{\varepsilon_{0}}a^{-4}}=\frac{\Omega_{b0}}{\Omega_{\gamma0}}a=\frac{\Omega_{b0}}{\Omega_{\gamma0}}\frac{1}{\left(1+z\right)}=\left.\frac{\Omega_{b}\left(z\right)}{\Omega_{\gamma}\left(z\right)}\right|_{\left\{ G,c\right\} =\text{const}}.\label{eq:Omega_b-over-Omega_g}
\end{equation}
We used Eqs. (\ref{eq:Omega_i}) and (\ref{eq:epsilon_m,r,L}) in
the second and third steps, respectively. Notice the cancelation of
$c\left(z\right)$; for this reason, (\ref{eq:Omega_b-over-Omega_g})
is the same result as in standard cosmology. Something similar occurs
in Eq. (\ref{eq:sond-horizon}). In fact, substituting (\ref{eq:c_s(z)})
and (\ref{eq:E(z)}) into (\ref{eq:r_d(z_d)}): \footnote{The dimensionless Hubble parameter $E\left(z\right)$ of Eq. (\ref{eq:E(z)})
could be taken as $E\left(z\right)\simeq\frac{c\left(z\right)}{c_{0}}\sqrt{\Omega_{m0}\left(1+z\right)^{3}+\Omega_{r0}\left(1+z\right)^{4}}$
for all practical effects of computing $r_{d}$ in (\ref{eq:r(z)-unchanged}).
This is because in the regime $[z_{d},\infty[$ of the integration
limits, the enegy contribution of $\Lambda$ is negligible in comparison
with the contribution of matter and radiation. The same goes for $E\left(z\right)$
in $r_{*}$.} 
\begin{equation}
r_{s}\left(z\right)=\frac{c_{0}}{H_{0}}\int_{z}^{\infty}\frac{1}{c_{0}}\frac{c\left(z^{\prime}\right)}{\sqrt{3\left[1+\frac{3}{4}\frac{\Omega_{b}\left(z^{\prime}\right)}{\Omega_{\gamma}\left(z^{\prime}\right)}\right]}}\frac{dz^{\prime}}{\frac{c\left(z^{\prime}\right)}{c_{0}}\sqrt{\Omega_{m0}\left(1+z^{\prime}\right)^{3}+\Omega_{r0}\left(1+z^{\prime}\right)^{4}+\Omega_{\Lambda}}}=\frac{c_{0}}{H_{0}}\int_{z}^{\infty}\left.\frac{c_{s}\left(z^{\prime}\right)}{c_{0}}\frac{dz^{\prime}}{E\left(z^{\prime}\right)}\right|_{\left\{ G,c\right\} =\text{const}},\label{eq:r(z)-unchanged}
\end{equation}
i.e. the cancelation of the function $c\left(z\right)$ within $c_{s}\left(z\right)$
with the function $c\left(z\right)$ within $E\left(z\right)$ renders
$r_{s}\left(z\right)$ unchanged with respect to the formula of standard
cosmology.

The last point of possible change in the sound horizon is regarding
the value of the redshift of baryon drag epoch $z_{d}$ (and/or the
value of the redshift of photon decoupling $z_{*}$). In fact, if
the value of $z$-drag changes in the context of our BD-like co-varying
$G$ and $c$ scenario, then $r_{d}=r_{s}\left(z_{d}\right)$ will
be modified (and similarly for $r_{*}$). Actually, we have demonstrated
in Section \ref{sec:Background-cosmology} that a varying $c$ modifies
the usual expression $T\sim a^{-1}\sim\left(1+z\right)$ to $T\sim a^{-1}c\left(a\right)\sim\left(1+z\right)c\left(z\right)$,
cf. Eq. (\ref{eq:T(a)}). Therefore, we would expect that $z_{d}$
does change in a general varying-$c$ cosmology; however, there is
a particular class of functions $c=c\left(z\right)$ that keeps the
value $z_{d}\simeq1060$ of standard cosmology unchanged; namely,
functions $c\left(z\right)$ for which the speed of light is equal
to $c_{0}$ today $\left(z=0\right)$ and is also equal to $c_{0}$
in the remote past $\left(z\gtrsim1000\right)$. We will choose parameterizations
for $c=c\left(z\right)$ that strictly follow this condition in Section
\ref{sec:Modelling-c(t)}. For this reason, $r_{d}$ (and $r_{*}$)
will be the same as is standard cosmology for all the varying-$c$
models studied in this paper.

Besides $r_{d}$, the other relavant quantity for modelling the baryon
accoustic oscillations is the average volume distance $d_{V}$ \cite{Percival2007}:
\begin{equation}
d_{V}\sim\left[r^{\parallel}\left(r^{\perp}\right)^{2}\right]^{1/3}.\label{eq:d_V(r_s)}
\end{equation}
Here $r^{\parallel}$ denotes the comoving radial distance from the
observer to the galaxy distribution of interest at redshift $z$ along
their line-of-sight. On the other hand, $r^{\perp}$ represents the
two comoving distances perpendicular to the line-of-sight. These comoving
distances are computed as follows: 
\begin{equation}
r^{\parallel}=\int\frac{c\left(z\right)}{H\left(z\right)}dz\Rightarrow r^{\parallel}\simeq\frac{c\left(z\right)}{H\left(z\right)}\int dz\simeq d_{h}\delta z\text{,}\label{eq:r_s-parallel}
\end{equation}
where we have assume that both $c\left(z\right)$ and $H\left(z\right)$
change very little in the square box containing the galaxy distribution
in the volume $d_{V}^{3}$. The quantity 
\begin{equation}
d_{h}\left(z\right)=\frac{c\left(z\right)}{H\left(z\right)}=\frac{1}{H_{0}}\frac{c\left(z\right)}{E\left(z\right)}\label{eq:d_h}
\end{equation}
is usually called the horizon distance (or Hubble distance). Also:
\begin{equation}
r^{\perp}=\frac{1}{a}d_{A}\left(z\right)\delta\theta\Rightarrow r^{\perp}=d_{p}\delta\theta\text{,}\label{eq:r_s-perp}
\end{equation}
where $\frac{d_{A}}{a}$ is the comoving angular-diameter distance---Eqs.
(\ref{eq:d_p(t_0)}) and (\ref{eq:d_A(z,c)-flat}); $d_{p}$ is the
proper distance; $\delta z$ and $\delta\theta$ are variations in
the directions $\parallel$ and $\perp$, respectively.\footnote{There are separate observations of BAO in the parallel direction to
the line-of-sight and in the orthogonal directions of the line-of-sight.
See, e.g. Table I in Ref. \cite{DESI2024}.}

Due to Eqs. (\ref{eq:r_s-parallel}) and (\ref{eq:r_s-perp}), the
relation (\ref{eq:d_V(r_s)}) reads: 
\begin{equation}
d_{V}\left(z\right)\equiv\left(zd_{h}d_{p}^{2}\right)^{1/3}.\label{eq:d_V(d_h,d_p)}
\end{equation}
The introduction of $z$ in the above definition of $d_{V}$ does
not compromise the scheme of distance definition in different cosmological
models, cf. Ref. \cite{Percival2007}. An explicit functional form
of $d_{V}$ in terms of the redshift is obtained by substituting Eqs.
(\ref{eq:d_p(z,c)}) and (\ref{eq:d_h}) into (\ref{eq:d_V(d_h,d_p)}):
\begin{equation}
d_{V}\left(z\right)=\frac{c_{0}}{H_{0}}\left[\frac{c\left(z\right)}{c_{0}}\frac{z}{E\left(z\right)}\right]^{1/3}\left[\int_{0}^{z}\frac{c\left(z^{\prime}\right)}{c_{0}}\frac{dz^{\prime}}{E\left(z^{\prime}\right)}\right]^{2/3}.\label{eq:d_V(z)}
\end{equation}

We argue below Eq. (\ref{eq:d_p(z,c)}) that $d_{p}$ is the same
as in standard cosmology; so is $d_{h}$. The reason is again the
cancelation of the factor $c\left(z\right)$: 
\begin{align}
d_{h} & =\frac{1}{H_{0}}\frac{c\left(z\right)}{E\left(z\right)}=\frac{1}{H_{0}}\frac{c\left(z\right)}{\frac{c\left(z\right)}{c_{0}}\sqrt{\Omega_{m0}\left(1+z\right)^{3}+\Omega_{r0}\left(1+z\right)^{4}+\Omega_{\Lambda}}}\nonumber \\
 & =\frac{1}{H_{0}}\frac{c_{0}}{\sqrt{\Omega_{m0}\left(1+z\right)^{3}+\Omega_{r0}\left(1+z\right)^{4}+\Omega_{\Lambda}}}=\left.\frac{1}{H_{0}}\frac{c_{0}}{E\left(z\right)}\right\vert _{\left\{ G,c\right\} =\text{const}}=\left.d_{h}\right\vert _{\left\{ G,c\right\} =\text{const}}.\label{eq:d_h-unchanged}
\end{align}
We have used Eq. (\ref{eq:E(z)}) for $E\left(z\right)$ after the
second equality. Since both $d_{h}$ and $d_{p}$ in the context of
our BD-like scenario are the same as $d_{h}$ and $d_{p}$ in the
context of standard cosmology, Eq. (\ref{eq:d_V(d_h,d_p)}) guarantees
that the average volume distance $d_{V}$ is also the same both for
BD-like cosmology and $\Lambda$CDM cosmology. The same conclusion
stems from Eq. (\ref{eq:d_V(z)}).

\section{Modelling the varying $c$ (and $G$) }\label{sec:Modelling-c(t)}

As stated previously (Section \ref{sec:Background-cosmology}), the
set of equations for our modified gravity cosmological model are not
complete in the sense that the co-varying couplings $c$ and $G$
can not be determined deductively; they enter as constitutive equations
whose free parameters are to be constrained from data. We will adopt
three distinct parameterizations for our varying speed of light. We
are interested in simplicity, phenomenological success and theoretical
naturalness%(no need for adding an ad hoc cut-off in the variation of $c$)
. Based on these criteria, we choose: (i) power-law parameterization
for $c\left(z\right)$, inspired by e.g. Refs.   \cite{Mendonca2021,Cuzinatto2023CPC,Cuzinatto2022}; (ii) exponential
parametrization by Gupta \cite{Gupta2020,Gupta2022,Gupta2022VCC,Gupta2023,Gupta2023Glob,Gupta2023JWST,Gupta2024,Gupta2024BAO}; (iii) continuous
parameterization of $c\left(z\right)$ with $c=c_{0}$ both today
and in the time before photon decoupling era. The analysis of these
models are the subject of the following subsections.

\subsection{Power-law parameterization}\label{subsec:Power-law-parameterization}

\begin{figure}[ht]
\begin{centering}
\includegraphics[width=0.60\textwidth]{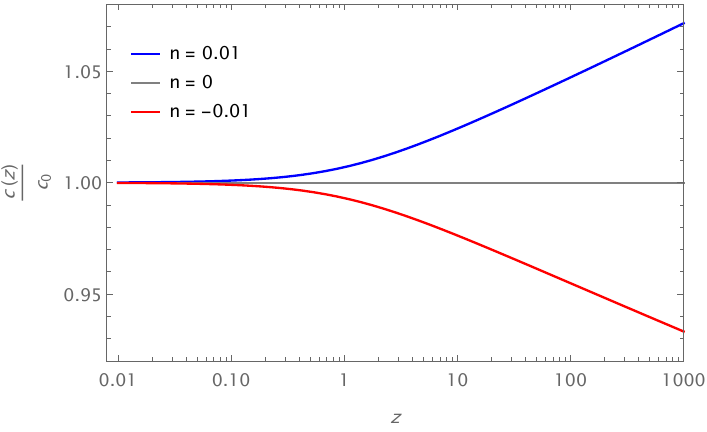}
\par\end{centering}
\caption{Redshift evolution of the speed of light $c\left(z\right)/c_{0}$
for the power-law parameterization with different values of the power-law
index $n$. }
\label{fig:c(z)-power-law}
\end{figure}

The first parameterization is one of the simplest approaches to modeling
a varying speed of light, the power-law parameterization---see e.g. \cite{Mendonca2021,Cuzinatto2023CPC}:
\begin{equation}
c\left(z\right)=c_{0}\left(1+z\right)^{n}.\label{eq:c(z)-power-law}
\end{equation}
This form is particularly useful for testing deviations from the standard
cosmological model, as it introduces only one additional parameter,
$n$, that controls the redshift dependence of the speed of light.
A positive (negative) value of $n$ implies that light traveled faster
(slower) in the past, while standard cosmology is recovered for $n=0$.
This behavior can be visualized in Figure \ref{fig:c(z)-power-law}.

The main advantage of the parameterization (\ref{eq:c(z)-power-law})
is its simplicity and generality. It allows one to easily check whether
current observational data favors any deviation from a constant $c_{0}$,
without relying on the underlying mechanism responsible for such variation.
However, the model requires the introduction of a redshift cut-off
$z_{\text{cut}}$, above which the speed of light returns to its standard
value $c_{0}$:
\begin{equation}
c\left(z\right)=\begin{cases}
c_{0}\left(1+z\right)^{n}, & z<z_{\text{cut}}\\
c_{0}, & z\geqslant z_{\text{cut}}
\end{cases}.\label{eq:c(z_cut)}
\end{equation}
This is necessary to avoid inconsistencies with well-established early-Universe
physics, such as Big Bang Nucleosynthesis and the Cosmic Microwave
Background, which are highly sensitive to the behavior of fundamental
constants at high redshifts ($z\gtrsim1000$). In our case, we are
interested in low-redshift phenomena, so we adopt cut-off values $z_{\text{cut}}=5$,
$10$, $50$, and $100$. We anticipate that the choice of $z_{\text{cut}}$
has a negligible impact on the parameter constraints, since the observational
data we use lie entirely below the cut-off; this expectation will
be checked in Section \ref{sec:Observations}.

\subsection{Gupta's parameterization}\label{subsec:Rajendra-parameterization}

The CCC model by Gupta has gained attention of the comunity lately
for its interesting phenomenological predictions---e.g. \cite{Gupta2023JWST,Gupta2024BAO};
it is based on the following parameterization of the varying speed
of light: 
\begin{equation}
c=c_{0}f\left(t\right)\qquad\text{with}\qquad f\left(t\right)=\exp\left[\alpha\left(t-t_{0}\right)\right].\label{eq:c(t)-Gupta}
\end{equation}
Note that $\alpha$ has dimensions of $\left(\text{time}\right)^{-1}$,
the same as the Hubble constant $H_{0}$.

The next steps include determining the function $c=c\left(z\right)$
in the context of our BD-Like VSL proposal. This parameterization
will require the adoption of a cut-off redshift, just like we did
for the power-law parameterization. However, similarly to the previous
case, we do not expect the particular value of $z_{\text{cut}}$ to
cause great impact for the results related to Gupta's parameterization.

Now, take Eqs. (\ref{eq:E(H)}) and (\ref{eq:E(a)}), 
\begin{equation}
E\left(a\right)=\frac{H}{H_{0}}=\frac{c\left(a\right)}{c_{0}}\sqrt{\Omega_{m0}a^{-3}+\Omega_{r0}a^{-4}+\Omega_{\Lambda}},\label{eq:E(a)-Gupta}
\end{equation}
for a negligible contribution from radiation. Therefore, 
\[
\frac{1}{H_{0}}\frac{1}{a}\frac{da}{dt}=\exp\left[\alpha\left(t-t_{0}\right)\right]\sqrt{\frac{\Omega_{m0}}{a^{3}}+\Omega_{\Lambda}},
\]
where $c=c\left(a\left(t\right)\right)$, i.e. $c=c\left(t\right)$
as in Eq. (\ref{eq:c(t)-Gupta}). The goal is to calculate $a\left(t\right)$
by integrating: 
\[
\int_{a_{0}}^{a}\frac{1}{a^{\prime}}\frac{da^{\prime}}{\sqrt{\frac{\Omega_{m0}}{a^{\prime3}}+\Omega_{\Lambda}}}=H_{0}\int_{t_{0}}^{t}\exp\left[\alpha\left(t^{\prime}-t_{0}\right)\right]dt^{\prime}.
\]
The analytical result is obtained without difficulty. It reads: 
\begin{equation}
\left(t-t_{0}\right)=\frac{1}{\alpha}\ln\left\{ 1+\frac{2}{3}\frac{1}{\Omega_{\Lambda}^{1/2}}\left(\frac{\alpha}{H_{0}}\right)\ln\left[\frac{\left(\frac{a}{a_{m\Lambda}}\right)^{3/2}+\sqrt{1+\left(\frac{a}{a_{m\Lambda}}\right)^{3}}}{\left(\frac{a_{0}}{a_{m\Lambda}}\right)^{3/2}+\sqrt{1+\left(\frac{a_{0}}{a_{m\Lambda}}\right)^{3}}}\right]\right\} ,\label{eq:t(a)-Gupta}
\end{equation}
where we have defined 
\begin{equation}
a_{m\Lambda}\equiv\left(\frac{\Omega_{m0}}{\Omega_{\Lambda}}\right)^{1/3}.\label{eq:amL-def}
\end{equation}

When Eq. (\ref{eq:E(a)}) is calculated at the present-day time $t_{0}$,
with $c\left(a_{0}\right)=c_{0}$, $\Omega_{r0}\simeq0$ and $a_{0}=1$,
it yields the constraint: 
\begin{equation}
1=\Omega_{m0}+\Omega_{\Lambda},\label{eq:Omegas-constraint-Gupta}
\end{equation}
so that $\Omega_{\Lambda}$ can be eliminated in favor of $\left(1-\Omega_{m0}\right)$
in Eq. (\ref{eq:t(a)-Gupta}).

Eq. (\ref{eq:z(a)}) relates $a$ to redshift $z$. With that, Eq.
(\ref{eq:t(a)-Gupta}) is written as: 
\begin{equation}
H_{0}\left(t-t_{0}\right)=\frac{1}{\beta}\ln\left\{ 1+\frac{2}{3}\frac{\beta}{\Omega_{\Lambda}^{1/2}}\ln\left[\frac{\left(\frac{a_{0}}{a_{m\Lambda}}\frac{1}{\left(1+z\right)}\right)^{3/2}+\sqrt{1+\left(\frac{a_{0}}{a_{m\Lambda}}\frac{1}{\left(1+z\right)}\right)^{3}}}{\left(\frac{a_{0}}{a_{m\Lambda}}\right)^{3/2}+\sqrt{1+\left(\frac{a_{0}}{a_{m\Lambda}}\right)^{3}}}\right]\right\} \label{eq:t(z,beta)}
\end{equation}
where 
\begin{equation}
\Omega_{\Lambda}=\left(1-\Omega_{m0}\right)\,,\qquad a_{m\Lambda}\equiv\left(\frac{\Omega_{m0}}{1-\Omega_{m0}}\right)^{1/3},\label{eq:Omega_L}
\end{equation}
and 
\begin{equation}
\beta\equiv\frac{\alpha}{H_{0}}\label{eq:beta}
\end{equation}
is the dimensionless parameter for data fitting. Eq. (\ref{eq:t(z,beta)})
provides the temporal dependence of the scale factor in Gupta's parameterization
in BD-like VSL---in the format $t=t\left(a\right)$.

\begin{figure}[ht]
\begin{centering}
\includegraphics[width=0.60\textwidth]{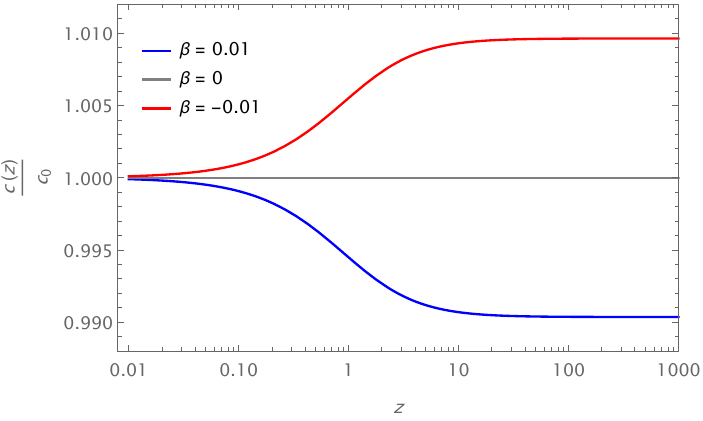}
\par\end{centering}
\caption{Redshift evolution of the speed of light $c\left(z\right)/c_{0}$
for the Gupta's parameterization for multiple values of $\beta$ with
$\Omega_{m0}=0.3$.}
\label{fig:c(z)-Gupta}
\end{figure}

Inserting (\ref{eq:t(z,beta)}) into (\ref{eq:c(t)-Gupta}): 
\begin{equation}
\frac{c\left(z\right)}{c_{0}}=1+\frac{2}{3}\frac{\beta}{\sqrt{1-\Omega_{m0}}}\ln\left[\frac{\sqrt{1-\Omega_{m0}}+\sqrt{\Omega_{m0}\left(1+z\right)^{3}+\left(1-\Omega_{m0}\right)}}{\left(1+z\right)^{3/2}\left(1+\sqrt{1-\Omega_{m0}}\right)}\right]\label{eq:c(z)-Gupta}
\end{equation}
This is a consistent result: taking $z=0$ in (\ref{eq:c(z)-Gupta})
gives $c\left(0\right)=c_{0}$, as expected. The behavior of $\frac{c\left(z\right)}{c_{0}}$
in this model can be visualized in Figure \ref{fig:c(z)-Gupta}.

Although the functional form of the power-law parameterization and
the functional form of Gupta's parametrization are distinct, the resulting
behavior of $c\left(z\right)$ is qualitatively similar over the redshift
range of interest---See Figs. \ref{fig:c(z)-power-law} and \ref{fig:c(z)-Gupta}.
However, the key difference lies in the growth behavior and asymptotic
structure of the two models. The power-law parametrization exhibits
a slow and continuous increase (or decrease) across all redshifts,
while Gupta's model features a more abrupt rise or fall that quickly
settles into an asymptotic constant value at high redshift.

\begin{figure}[ht]
\begin{centering}
(a)\includegraphics[width=0.47\textwidth]{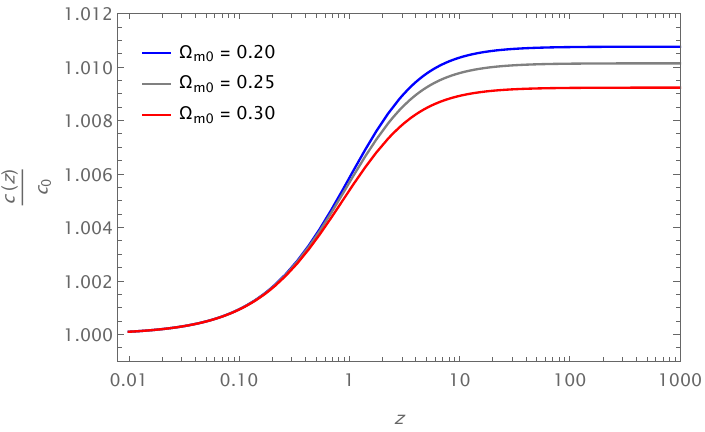}\vspace{0.5cm} (b)\includegraphics[width=0.47\textwidth]{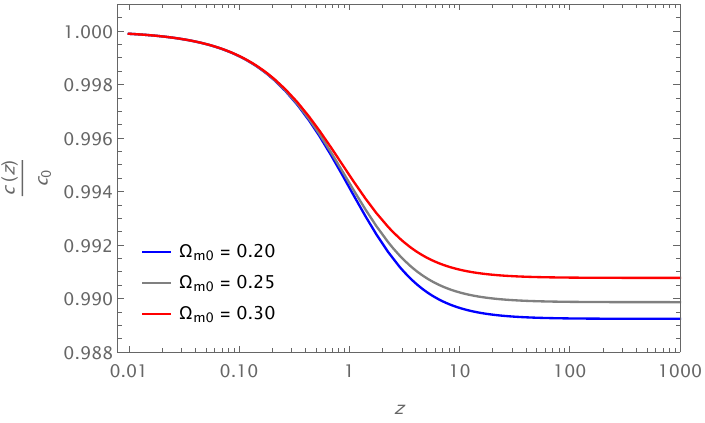}
\par\end{centering}
\caption{Redshift evolution of the speed of light $c\left(z\right)/c_{0}$
for Gupta's parameterization and different values of $\Omega_{m0}$
with (a) $\beta=-0.01$, and (b) with $\beta=0.01$.}
\label{fig:c(z)-Gupta-beta-neg-pos}
\end{figure}

In addition, the influence of the matter density parameter $\Omega_{m0}$
on the variation of $c\left(z\right)$ is found to be minimal in Gupta's
CCC parameterization. The shape of the curves remains structurally
the same regardless of the value of $\Omega_{m0}$; only the asymptotic
value of $c\left(z\right)$ is affected. For negative $\beta$, lower
values of $\Omega_{m0}$ lead to higher asymptotic values of $c\left(z\right)$,
as shown in Figure \ref{fig:c(z)-Gupta-beta-neg-pos}(a). For positive
$\beta$, smaller $\Omega_{m0}$ results in lower asymptotic values
of $c\left(z\right)$, as demonstrated in Figure \ref{fig:c(z)-Gupta-beta-neg-pos}(b).

\subsection{Continuous parameterization}\label{subsec:Continuous-parameterization}

To address the limitation of the previous parameterizations---namely,
the need to impose an ad hoc redshift cut-off to avoid conflicts with
early-Universe physics---we consider the continuous parameterization:
\begin{equation}
c\left(z\right)=c_{0}\left\{ 1-\exp\left[-\left(1+\alpha\right)z\right]+\exp\left(-\alpha z\right)\right\} ^{n},\label{eq:c(z)-cont}
\end{equation}
where $\alpha>0$. This model naturally resolves the issue of a cut-off
by smoothly interpolating between the present-day value $c\left(0\right)=c_{0}$
and an asymptotic early-time value $\lim_{z\rightarrow\infty}c\left(z\right)=c_{0}$,
thereby eliminating the need to explicitly define a transition redshift
where the variation of $c\left(z\right)$ stops. Moreover, in the
low-redshift regime where $\alpha z\ll1$, this parametrization approximates
the power-law form: 
\begin{equation}
c\left(z\right)\simeq c_{0}\left\{ 1-\left[1-\left(1+\alpha\right)z\right]+\left(1-\alpha z\right)\right\} ^{n}=c_{0}\left(1+z\right)^{n}.\label{eq:c(z)-cont-approx}
\end{equation}

\begin{figure}
\begin{centering}
(a)\includegraphics[width=0.48\textwidth]{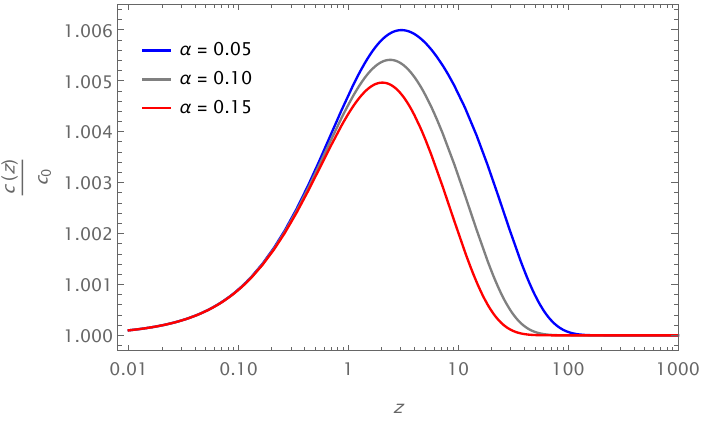}\vspace{0.5cm}(b)\includegraphics[width=0.48\textwidth]{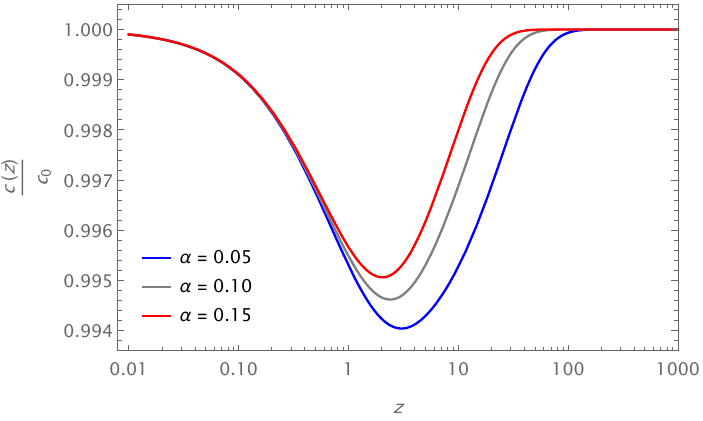}
\par\end{centering}
\caption{(a) Redshift evolution of the speed of light $c\left(z\right)/c_{0}$
for the continuous parameterization, multiple values of $\alpha$,
with power-law index $n=0.01$ in part (a) and $n=-0.01$ in part
(b).}
\label{fig:c(z)-cont-n-pos-neg}
\end{figure}

The parameterization in Eq. (\ref{eq:c(z)-cont}) follows the same
logic as the power-law model: a positive (negative) value of $n$
implies that light traveled faster (slower) in the past, while standard
cosmology with a constant speed of light is recovered for $n=0$.
The additional parameter $\alpha$ regulates how quickly the variation
in $c\left(z\right)$ transitions back to its present value $c_{0}$.
Larger (smaller) values of $\alpha$ imply that the deviation from
the standard behavior was less (more) pronounced. This behavior can
be visualized in Figure \ref{fig:c(z)-cont-n-pos-neg}.

In the limit $n\rightarrow0$, the variation becomes negligible, and
the model reduces to General Relativity. As both $n$ and $\alpha$ suppress the deviation from $c_{0}$, a potential degeneracy between them will arise in data analysis. To
avoid this degeneracy, we fix $\alpha=0.01$, $0.1$ and $1$. 

\section{Constraining the varying $c$ and $G$ cosmology }\label{sec:Observations}

\subsection{Datasets}\label{subsec:Data-sets}

The datasets used to constrain our BD-like co-varying $G$
and $c$ model are the following:
\begin{enumerate}
\item SNe Ia data from the Pantheon$+$ set \cite{Pantheon+} and Union2.1
set \cite{Union2.1}. The relevant observational quantity for these
datasets is the luminosity distance, given by Eqs. (\ref{eq:d_p(t_0,z)})
and (\ref{eq:d_L(z,c)-flat}). Since SN Ia data correspond to relatively
low redshifts, $z\in\left[0,3\right]$, and $\Omega_{r0}\sim10^{-5}$,
we can approximate:
\begin{equation}
d_{L}\left(z\right)=\left(1+z\right)\frac{c\left(z\right)}{H_{0}}\int_{0}^{z}\frac{dz^{\prime}}{\sqrt{\Omega_{m0}\left(1+z^{\prime}\right)^{3}+\Omega_{\Lambda}}},\label{eq:d_L(z)}
\end{equation}
without any negative impact in the constraining power. The likelihood
function $\mathcal{L}$ used to fit cosmological models to the SN
Ia data is given by: 
\begin{equation}
\mathcal{L}\propto\exp\left(-\frac{1}{2}\Delta\mu^{T}C^{-1}\Delta\mu\right),\label{eq:SN_Likelihood}
\end{equation}
where $\Delta\mu = \mu_{obs}-\mu_{theo}$ is a vector with the difference between the the observed (obs) and theoretical (theo) distance modulus. $C$ is the full covariance matrix \cite{GelmanRubin}; this
matrix accounts for both statistical (stat) and systematic (syst)
uncertainties; it also incorporates expected correlations among the
light curves of the supernovae in the sample. Specifically, we write
$C=C_{\text{stat}}+C_{\text{syst}}$. This procedure is applied for both the Pantheon+ and Union2.1 datasets. The details of the Pantheon+ dataset can be found
in Brout et al. {\cite{Pantheon+}}, while those of Union2.1 can be found in Suzuki et al. {\cite{Union2.1}}.
\item BAO data from DESI Collaboration DR2 \cite{DESI2025EDE}. The relevant
observational quantity for this datasets is the average volume distance
expressed in a dimensionless form---see Eq. (\ref{eq:d_V(z)}): 
\begin{equation}
D_{V}\left(z\right)\equiv\frac{d_{V}\left(z\right)}{r_{d}}=\frac{c_{0}}{r_{d}H_{0}}\left[\frac{c\left(z\right)}{c_{0}}\frac{z}{E\left(z\right)}\right]^{1/3}\left[\int_{0}^{z}\frac{c\left(z^{\prime}\right)}{c_{0}}\frac{dz^{\prime}}{E\left(z^{\prime}\right)}\right]^{2/3}.\text{ }\label{eq:D_V(r_d,z)}
\end{equation}
In principle, BAO data carry information both on the more recent universe
through the average volume distance $d_{V}$ and on the pre-recombination
universe through the sound horizon computed at drag epoch, $r_{d}=r\left(z_{d}\right)$
cf. Eq. (\ref{eq:r_d(z_d)}). However, BAO data alone cannot independently
determine the value of the Hubble constant $H_{0}$. This is because
BAO measurements are directly related to the product $r_{d}H_{0}$.
Due to this limitation, it is useful to define:
\begin{equation}
K\left(z\right)\equiv\frac{r\left(z\right)H_{0}}{c_{0}},\label{eq:K(z)}
\end{equation}
so that $K_{d}=\frac{r_{d}H_{0}}{c_{0}}$, where $K_{d}\equiv K\left(z_{d}\right)$. Moreover, analogous to the
SNe Ia case, the DESI sets inform about the recent universe since
BAO redshifts are relatively low, $z\in\left[0,3\right]$ and the
constraint $\Omega_{m0}+\Omega_{r0}+\Omega_{\Lambda}=1$ (coming from
Eq. (\ref{eq:E(z)}) at $z=0$) might as well be approximated by $\Omega_{m0}+\Omega_{\Lambda}\simeq1$
since $\Omega_{r0}=\Omega_{\gamma0}+\Omega_{\nu0}\simeq8.4\times10^{-5}$.
So, we have: 
\begin{equation}
D_{V}\left(z\right)=\frac{1}{K_{d}}\left[\frac{c\left(z\right)}{c_{0}}\frac{z}{\sqrt{\Omega_{m0}\left(1+z^{\prime}\right)^{3}+\Omega_{\Lambda}}}\right]^{1/3}\left[\int_{0}^{z}\frac{c\left(z^{\prime}\right)}{c_{0}}\frac{dz^{\prime}}{\sqrt{\Omega_{m0}\left(1+z^{\prime}\right)^{3}+\Omega_{\Lambda}}}\right]^{2/3}.\label{eq:D_V(K_d,z)}
\end{equation}
The likelihood function used to fit cosmological models to the BAO
data is given by: 
\begin{equation}
\mathcal{L}\propto\exp\left\{ -\frac{1}{2}\left[\frac{D_{V}^{\text{obs}}\left(z\right)-D_{V}^{\text{theo}}\left(z\right)}{\sigma\left(z\right)}\right]^{2}\right\} ,\label{eq:BAO_Likelihood}
\end{equation}
where $D_{V}^{\text{obs}}\left(z\right)-D_{V}^{\text{theo}}\left(z\right)$
is the difference between the observed (obs) and theoretical (theo)
dimensionless average volume distance. $\sigma\left(z\right)$ is
the associated uncertainty.
\item Angular accoustic scale $\theta_{\ast}$ obtained from the position
of the first peak in CMB power spectrum \cite{Planck2018VI}. This
is the sole data constraining the cosmological model parameter through
Eqs. (\ref{eq:theta_star}) and (\ref{eq:d_p(z_star)}):
\begin{equation}
\theta_{\ast}=\frac{r_{\ast}H_{0}(1+z_{*})}{c_{0}}\left[\int_{0}^{z_{\ast}}\frac{c\left(z\right)}{c_{0}}\frac{dz}{E\left(z\right)}\right]^{-1}.\label{eq:theta_star(E(z))}
\end{equation}
The literature shows that data constraining using $\theta_{\ast}$
is robust with respect to different cosmological models, even taking
into account the fact that this parameter stems from the perturbation
theory applied to each of these distinct models \cite{Verde2022}.
This is one of the motivations for us to use this dataset to constrain
our VSL model. We implement a Gaussian external prior on the quantity
$100\theta_{\ast}$ with mean $1.04100$ and variance $\left(0.00030\right)^{2}$
\cite{Planck2018VI}. To align it more closely with the BAO case, it
is convenient to define: $K_{\ast}\equiv\frac{r_{\ast}H_{0}}{c_{0}}$,
with $K_{\ast}\equiv K\left(z_{*}\right)$,
in accordance with Eq. (\ref{eq:K(z)}). It is noteworthy that the
values of $z_{\ast}\simeq1090$ and $z_{d}\simeq1060$ are close \cite{Planck2018VI},
but not the same. As a result $r_{\ast}\neq r_{d}$, and consequently
$K_{\ast}$ $\neq$ $K_{d}$. However, due to the proximity of these
redshifts, we can treat one as a small correction to the other. This
allows us to consider a linear relation between them. Thus, we expand
$K\left(z\right)$ around $z=z_{d}$ to first order using a Taylor
expansion:
\begin{equation}
K\left(z\right)\simeq K_{d}+\left.\frac{dK}{dz}\right\vert _{z=z_{d}}\left(z-z_{d}\right),\label{eq:K(z)-Taylor-z_d}
\end{equation}
where the derivative term accounts for the redshift sensitivity of
the sound horizon. Therefore, we have: 
\begin{equation}
K_{\ast}\simeq K_{d}-6\times10^{-4}.\label{eq:K_star(K_d)}
\end{equation}
\end{enumerate}

\subsection{VSL model constraining }\label{subsec:VSL-model-constraining}

In this subsection, we perform a comparative analysis between the three
different parameterizations  specified in Section \ref{sec:Modelling-c(t)}
for our BD-like VSL framework. 

The equations developed in Subsection \ref{subsec:Data-sets} were
implemented in Python, with flat priors applied to all free parameters,
as shown in Table \ref{tab:Priors}. Cosmological parameter sampling
was performed using the Monte Carlo Markov Chain (MCMC) method via
the emcee library \cite{emcee}. A convergence diagnostic based on
the Gelman--Rubin criterion \cite{GelmanRubin} was applied, with
a threshold set to $R-1\lesssim0.01$ \cite{GelmanRubin}. Additionally,
GetDist \cite{GetDist} was used for post-processing and visualization
of the MCMC outputs, allowing for the generation of contour plots
and marginalized distributions of the cosmological parameters at $1\sigma$
and $2\sigma$ confidence levels.

\begin{table}[htbp]
\begin{centering}
\begin{tabular}{|c|c|c|c|c|c|}
\hline 
Parameter & $\Omega_{m}$ & $K_{d}$ & $H_{0}$ & $n$ & $\beta$\tabularnewline
\hline 
\hline 
Prior & $\left[0,1\right]$ & $\left[0.020,0.040\right]$ & $\left[40,100\right]$ & $\left[-1,1\right]$ & $\left[-1,1\right]$\tabularnewline
\hline 
\end{tabular}
\par\end{centering}
\caption{Flat priors for all free parameters in $c\left(z\right)$ parameterizations
used in our the BD-like VSL model.}
\label{tab:Priors}
\end{table}

\begin{figure}[ht]
\begin{centering}
\includegraphics[scale=0.8]{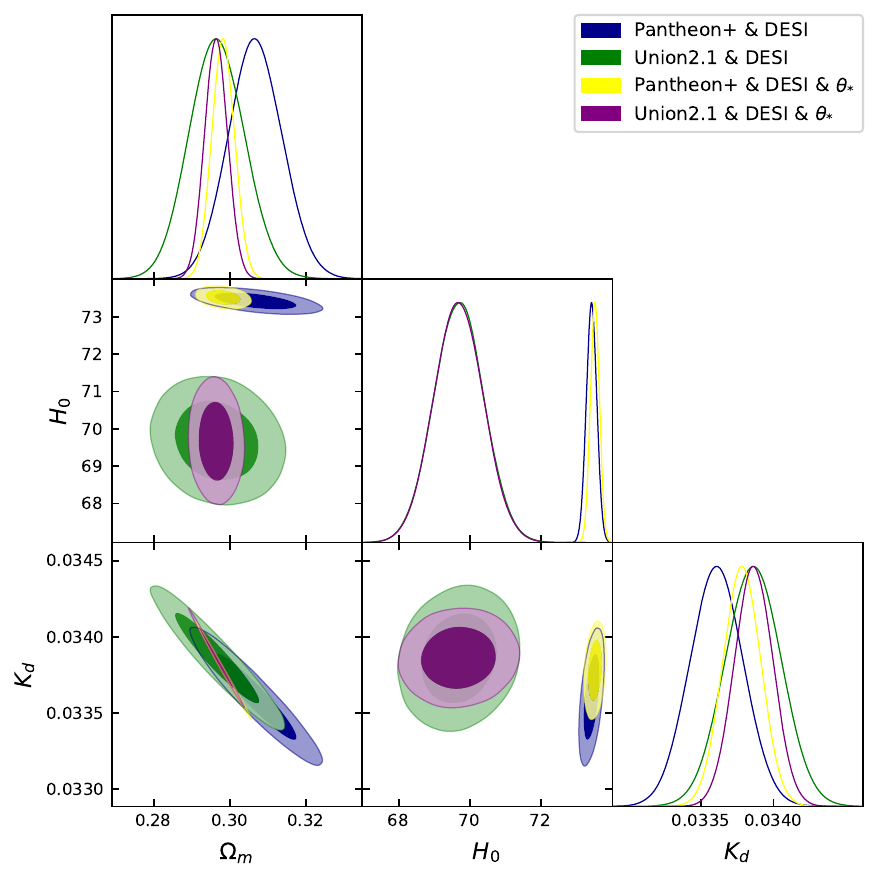}
\par\end{centering}
\caption{[$\Lambda$CDM model] 68\% and 95\% confidence level posterior distributions and contour
plots of the $\Lambda$CDM model for the parameters $\Omega_{m}$, $H_{0}$, $K_{d}$,
using the fit to the datasets Pantheon$+$ \cite{Pantheon+}, Union2.1 \cite{Union2.1}, DESI \cite{DESI2025EDE} and $\theta_{\ast}$ \cite{Planck2018VI}.}
\label{fig:contour-plot-LCDM}
\end{figure}

Before constraining the three parameterizations for the varying speed
of light in the context of our Brans-Dicke-like model, it is instructive
to fit the standard $\Lambda$CDM model using the same combination
of datasets to be employed in our VSL-parameterizations. As established in the literature
\cite{Valentino2021,Knox2020,Freedman2021}, there exists a significant discrepancy
in the inferred values of $H_{0}$ from different combinations of datasets.
This discrepancy is clearly illustrated in the contour plots of Fig. \ref{fig:contour-plot-LCDM} and in the top sector of Table \ref{tab:parameters-BD-like-VSL}. In particular, the ``Union2.1 \& DESI'' combination yields a lower value of the Hubble constant compared to ``Pantheon+ \& DESI''. It is important to note that the supernova catalogs considered alone do not exhibit any significant tension. The situation changes once the BAO data are
included, which introduces a substantial difference in the inferred $H_{0}$
as a consequence of the degeneracy associated with the sound-horizon scale
\cite{Pogosian2020,Escudero2025,Jedamzik2020}. In this context, the lower precision of the Union2.1 sample shifts the inferred Hubble constant towards smaller values, while the Pantheon$+$ sample, owing to its reduced uncertainties, favors higher values for $H_0$.

%That this discrepancy in $H_{0}$
%originates from the statistical structure of the datasets rather than from
%any assumption of a varying speed of light. The objective of the present
%work is instead to investigate how the inferred variation in $H_{0}$ can be
%reinterpreted within the VSL-parameterizations, a discussion which will be
%addressed in the following subsection.

\begin{figure}[ht]
\begin{centering}
\includegraphics[scale=0.8]{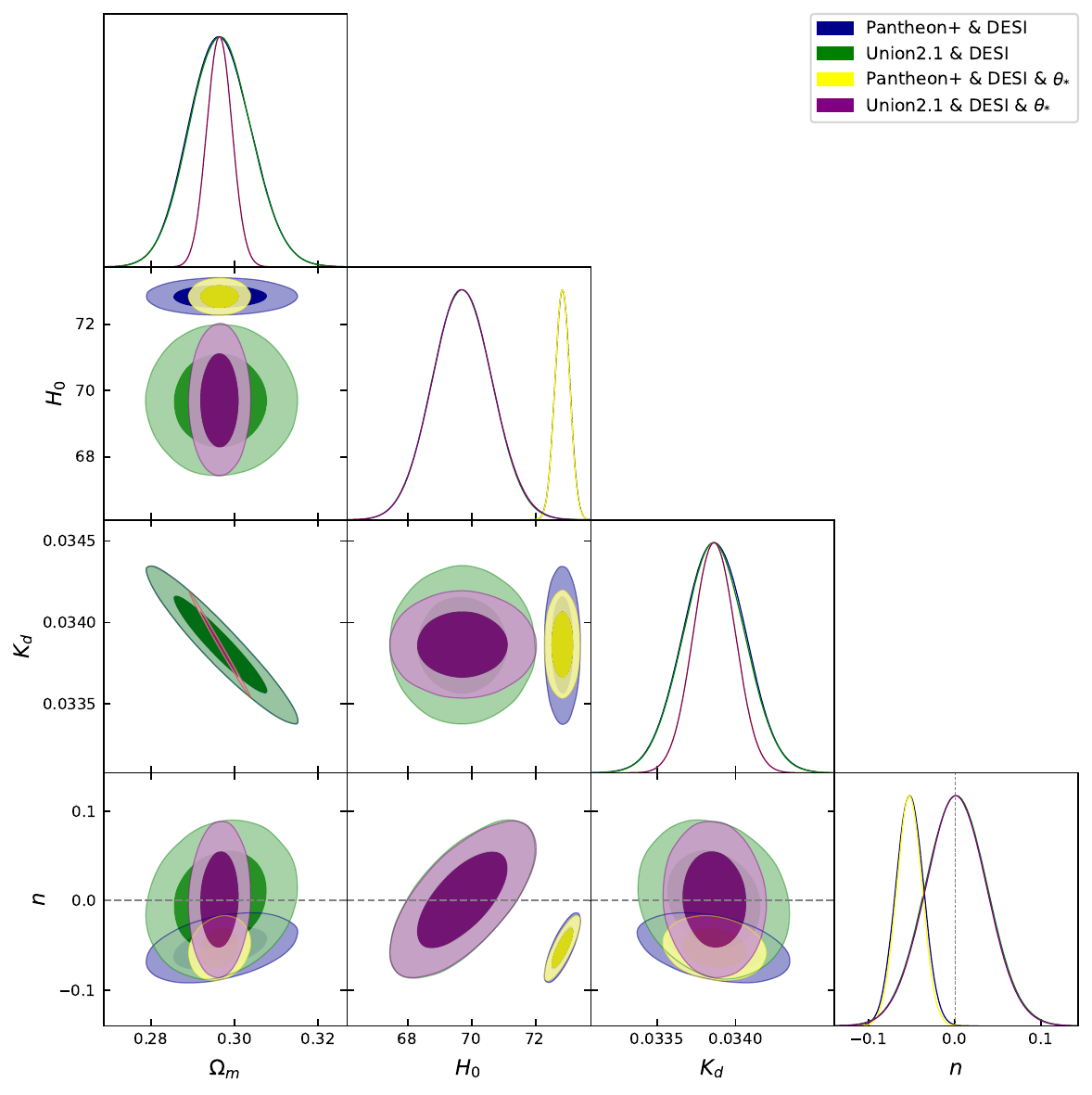}
\par\end{centering}
\caption{{[}Power-law parameterization{]} 68\% and 95\% confidence level posterior
distributions and contour plots of the VSL model with the power-law
parameterization for the parameters $\Omega_{m}$, $H_{0}$, $K_{d}$, $n$, using
the fit to the datasets redPantheon$+$ \cite{Pantheon+}, Union2.1 \cite{Union2.1}, DESI \cite{DESI2025EDE} and $\theta_{\ast}$ \cite{Planck2018VI}.}
\label{fig:contour-plot-power-law}
\end{figure}

\begin{figure}[ht]
\begin{centering}
\includegraphics[scale=0.8]{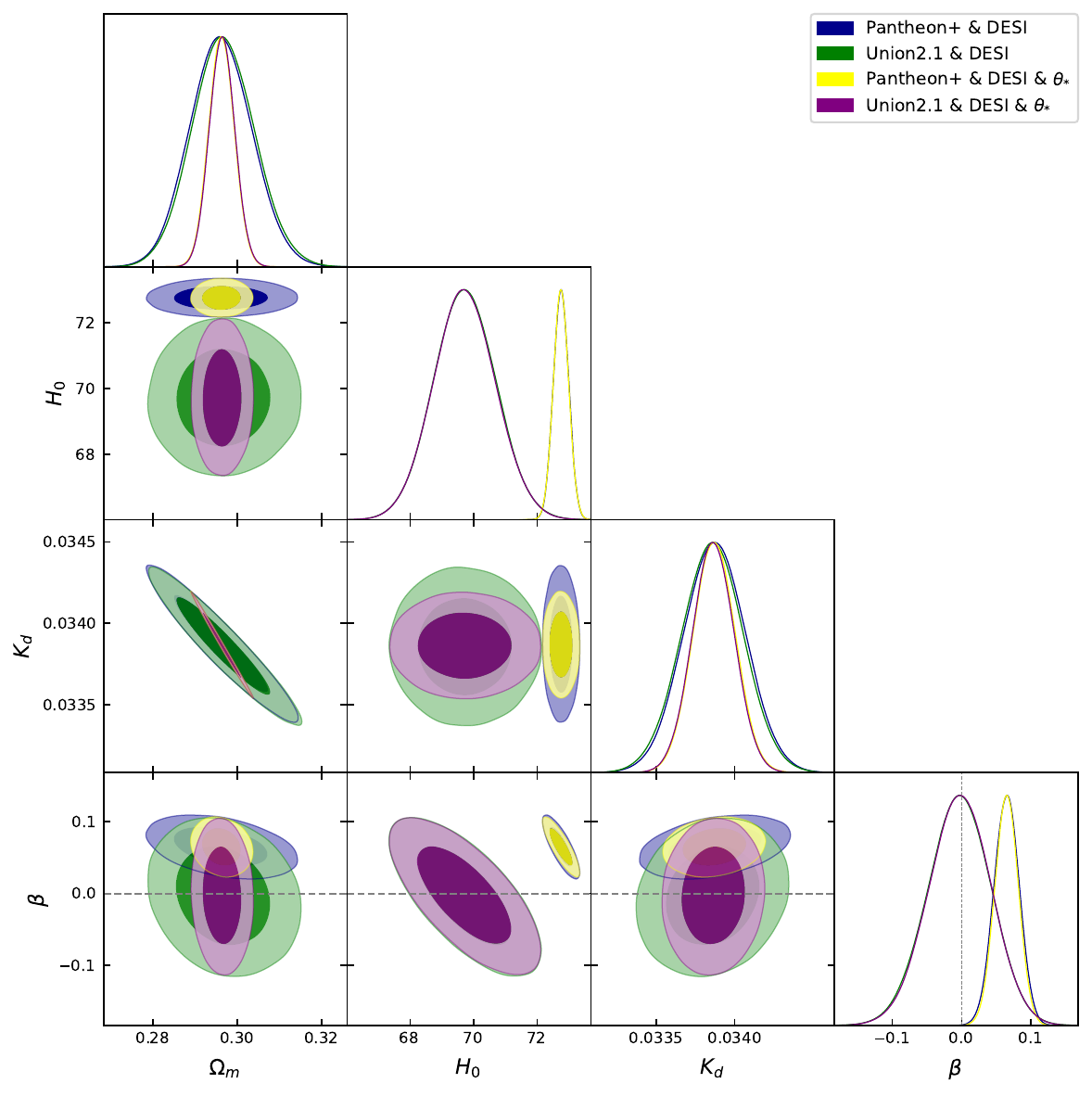}
\par\end{centering}
\caption{{[}Gupta's parameterization{]} 68\% and 95\% confidence level posterior
distributions and contour plots of the VSL model with Gupta's parameterization
for the parameters $\Omega_{m}$, $H_{0}$, $K_{d}$, $\beta$, using the fit
to the datasets Pantheon$+$ \cite{Pantheon+}, Union2.1 \cite{Union2.1}, DESI \cite{DESI2025EDE} and $\theta_*$ \cite{Planck2018VI}.}
\label{fig:contour-plot-Rajendra}
\end{figure}

\begin{figure}[ht]
\begin{centering}
\includegraphics[scale=0.8]{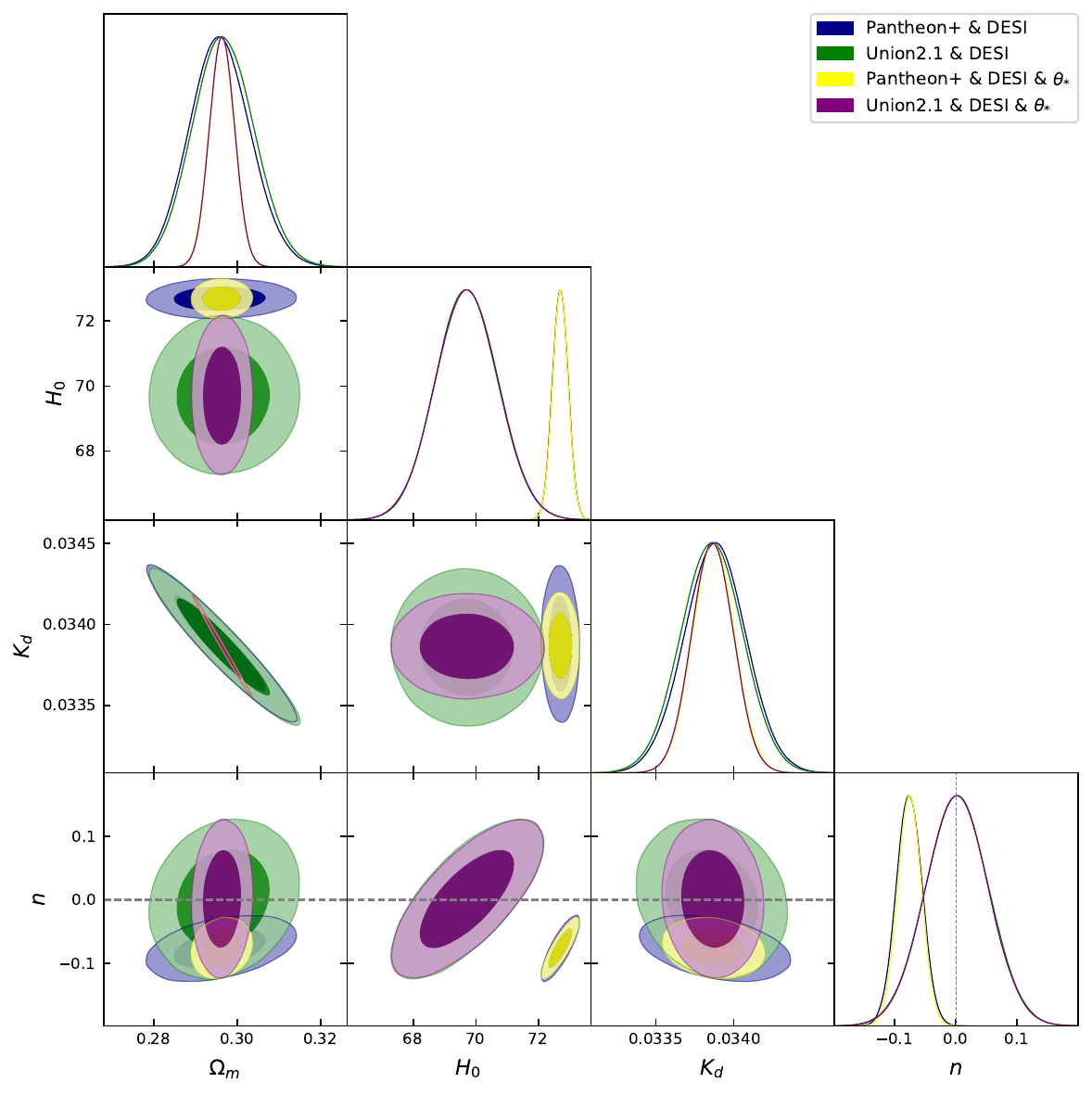}
\par\end{centering}
\caption{{[}Continuous parameterization{]} 68\% and 95\% confidence level posterior
distributions and contour plots of the VSL model with the continuous
parameterization for the parameters $\Omega_{m}$, $H_{0}$, $K_{d}$, $n$, using
the fit to the datasets Pantheon$+$ \cite{Pantheon+}, Union2.1 \cite{Union2.1}, DESI \cite{DESI2025EDE} and $\theta_*$ \cite{Planck2018VI}.}
\label{fig:contour-plot-continuous}
\end{figure}

We now turn to the VSL parameterizations. The main results of the data fitting process are summarized in Figs. \ref{fig:contour-plot-power-law}, \ref{fig:contour-plot-Rajendra},
\ref{fig:contour-plot-continuous}, and Table \ref{tab:parameters-BD-like-VSL}.
In the following, we discuss the conclusions that are extracted from these plots
and table.

Concerning the parameters $\Omega_{m}$ and $K_{d}$, we see that
for all VSL--parameterizations (power-law, Gupta's and continuous),
the values of the modes (central tendencies) do not change significantly
for all combinations of datasets (``Pantheon+ \& DESI'', ``Union2.1 \& DESI'', ``Pantheon+ \& DESI \& $\theta_*$'', ``Union2.1 \& DESI \& $\theta_*$''). For these parameters, the use
of the $\theta_*$ data has a significant role: to decrease
the dispersion of the distributions of both quantities $K$ and $\Omega_{m}$.
We also see that both the modes and the dispersions seem to be insensitive
to the use of the datasets from Union or Pantheon collaborations.
In particular, the physics underlying the acoustic oscillations, both
for matter (BAO) and CMB, is the main responsible for constraining $\Omega_{m}$---this
is why their distributions are insensitive to the use of SN Ia data;
the central tendencies of the $\Omega_{m}$ distributions are consistent,
which indicates the consistency both in the matter and in the electromagnetic
spectra.

On the other hand, the values of the Hubble constant, $H_{0}$, are
strongly dependent on the SN Ia dataset. The central tendency values
obtained with the data from Pantheon$+$ are about $72.7\,{\rm km/s\cdot Mpc}$.
The Union2.1 data, by their turn, present
higher values of uncertainty, so that the weighted value of $H_{0}$
has a significant contribution from the BAO data, which present smaller
uncertainties. This pushes the central tendency of $H_{0}$ to values
about $69.7\,{\rm km/s\cdot Mpc}$ when using the Union2.1 dataset.

With respect to the values of the free parameters of the models ($n$
or $\beta$), we observe that they are strongly dependent on the values
of $H_{0}$. In the case of Union datasets (``Union2.1 \& DESI'', ``Union2.1
\& DESI \& $\theta_*$''), where the values of $H_{0}$ are
consistent with the physics of the early Universe (CMB physics), the
values of the free parameter do not indicate any variation of the
speed of light---this is true for all the three VSL-parameterizations
presented here. This result is linked to the fact that the baryonic
acoustic oscillations result in values of $H_{0}$ consistent with
the values of CMB when the degeneracy of $K_{d}$ is broken (recall
that it is related to the Hubble constant and the speed of sound). 

The data from Pantheon$+$ (``Pantheon+ \& DESI'', ``Pantheon+
\& DESI \& $\theta_*$''), on the other hand, present higher
values of $H_{0}$ and deliver a probability distribution that indicates
that $c$ varies with respect to time with more than 99\% of confidence level.
Interestingly, the three models present a consistent qualitative behavior
for the speed of light with $c(z)<c_{0}$. We see that the resulting
distributions for $n$ or $\beta$ are indifferent to the use or not
of the $\theta_*$ dataset. These distributions present
almost negligible correlations with $\Omega_{m}$ and $K_{d}$, but,
on the contrary, they are strongly correlated to $H_{0}$.

\begin{table}[ht]
\centering
\caption{Results for Model Constraining}
\label{tab:parameters-BD-like-VSL}
\begin{threeparttable}
\begin{tabular}{lcccc}
\toprule
& $\Omega_m$ & $K_d$ & $H_0 [Km\,s^{-1}Mpc^{-1}]$ & Extra \\
\midrule

\multicolumn{5}{l}{\textbf{$\Lambda$CDM}} \\

Pantheon+ \& DESI & $0.3067 \pm 0.0071$ & $0.03361 \pm 0.00019$ & $73.43 \pm 0.15$ & -- \\
Union2.1 \& DESI & $0.2966 \pm 0.0073$ & $0.03386 \pm 0.00020$ & $69.68 \pm 0.70$ & -- \\
Pantheon+ \& DESI \& $\theta_*$ & $0.2982 \pm 0.0030$ & $0.03378 \pm 0.00013$ & $73.52 \pm 0.13$ & -- \\
Union2.1 \& DESI \& $\theta_*$ & $0.2964 \pm 0.0030$ & $0.03386 \pm 0.00013$ & $69.68 \pm 0.70$ & -- \\

\midrule
\multicolumn{5}{l}{\textbf{Power-law parameterization}} \\

Pantheon+ \& DESI & $0.2965 \pm 0.0074$ & $0.03386 \pm 0.00020$ & $72.84 \pm 0.23$ & $n = -0.053 \pm 0.016$ \\
Union2.1 \& DESI & $0.2966 \pm 0.0074$ & $0.03386 \pm 0.00020$ & $69.70 \pm 0.93$ & $n = 0.001 \pm 0.036$ \\
Pantheon+ \& DESI \& $\theta_*$ & $0.2963 \pm 0.0030$ & $0.03386 \pm 0.00013$ & $72.84 \pm 0.23$ & $n = -0.053 \pm 0.016$ \\
Union2.1 \& DESI \& $\theta_*$ & $0.2963 \pm 0.0034$ & $0.03386 \pm 0.00013$ & $69.71 \pm 0.93$ & $n = 0.001 \pm 0.036$ \\

\midrule
\multicolumn{5}{l}{\textbf{Gupta’s parameterization}} \\

Pantheon+ \& DESI & $0.2961 \pm 0.0074$ & $0.03387 \pm 0.00020$ & $72.75 \pm 0.24$ & $\beta = 0.065 \pm 0.018$ \\
Union2.1 \& DESI & $0.2967 \pm 0.0074$ & $0.03386 \pm 0.00020$ & $69.73 \pm 0.98$ & $\beta = -0.004 \pm 0.045$ \\
Pantheon+ \& DESI \& $\theta_*$ & $0.2962 \pm 0.0030$ & $0.03387 \pm 0.00013$ & $72.75 \pm 0.24$ & $\beta = 0.065 \pm 0.017$ \\
Union2.1 \& DESI \& $\theta_*$ & $0.2964 \pm 0.0030$ & $0.03386 \pm 0.00013$ & $69.72 \pm 0.97$ & $\beta = -0.003 \pm 0.045$ \\

\midrule
\multicolumn{5}{l}{\textbf{Continuous parameterization ($\alpha=0.01$)}} \\

Pantheon+ \& DESI & $0.2958 \pm 0.0074$ & $0.03388 \pm 0.00020$ & $72.69 \pm 0.25$ & $n = -0.076 \pm 0.021$ \\
Union2.1 \& DESI & $0.2967 \pm 0.0074$ & $0.03386 \pm 0.00020$ & $69.71 \pm 0.99$ & $n = 0.002 \pm 0.051$ \\
Pantheon+ \& DESI \& $\theta_*$ & $0.2962 \pm 0.0030$ & $0.03387 \pm 0.00013$ & $72.69 \pm 0.25$ & $n = -0.076 \pm 0.020$ \\
Union2.1 \& DESI \& $\theta_*$ & $0.2963 \pm 0.0030$ & $0.03386 \pm 0.00013$ & $69.70 \pm 1.00$ & $n = 0.002 \pm 0.051$ \\

\bottomrule
\end{tabular}

\begin{tablenotes}
\footnotesize
\item Notes: Summary of marginalized parameter constraints for Pantheon+, Union2.1 and DESI datasets. The mean and 68\% confidence limits are provided for each cosmological parameter. $\theta_*$ denotes the inclusion of the angular scale of the sound horizon from CMB. An additional column “Extra” is included, which lists the extra parameter introduced by each parametrization with respect to the standard $\Lambda$CDM model.
\end{tablenotes}
\end{threeparttable}
\end{table}

In summary, we can split the four parameters that we fit in each evaluation
in two distinct sets, namely: $\{\Omega_{m},K_{d}\}$ and $\left\{ H_{0},n\left(\text{or }\beta\right)\right\} $.
The parameters of the first set are mutually correlated %between themselves
and have strong influence from BAO and $\theta_*$ data.
Once the modeling of these latter quantities does not depend on the
variation $c\left(z\right)$, the two parameters are essentially independent
of the values of $n\left(\text{or }\beta\right)$. The parameters
of the second set, $\left\{ H_{0},n\left(\text{or }\beta\right)\right\} $,
are mutually correlated %between themselves 
and are independent of the other
two parameters. This correlation can somehow be expected since the
variation of $c$, $c\left(z\right)$, modifies the luminosity distance
$d_{L}$ which also depend on the values of $H_{0}$. Finally, we
point out that the ``Union2.1 \& DESI'' dataset suggest no variation of
the speed of light, while the ``Pantheon+ \& DESI'' data strongly
favors a variable speed of light with more than 3$\sigma$ confidence
level for all the three parameterizations analyzed here---more specifically,
3.3$\sigma$ for the power-law type, 3.8$\sigma$ for both Gupta's parameterization
and the continuous parameterization.

\subsection{Further discussions}
\paragraph*{Pantheon+ and Union 2.1:}
In this work, we consider two Type Ia supernova datasets, Pantheon$+$ and Union2.1.  Within our Brans--Dicke-like VSL framework, the only departure from the standard  $\Lambda$CDM model arises in the luminosity distance $d_L$. This is clearly illustrated in Eq.~(\ref{eq:d_L(z,c)-flat}): once the condition $c(z)=c_0$ is imposed, the standard scenario is immediately recovered. Since Type Ia supernova observations are directly sensitive to $d_L$, they constitute the unique late-time cosmological probe through which deviations from the standard scenario can manifest in our model.

When considering determinations of the Hubble constant based exclusively on late-Universe observations, different supernova compilations are known to be highly consistent with each other. In particular, late-time measurements of $H_0$ typically agree within $1\sigma$, as summarized, for instance, in Table~1 of Ref.~\cite{uddin2023}.

Motivated by this strong statistical compatibility, we focus on two representative SN Ia datasets. Pantheon$+$ is chosen because it constitutes one of the most comprehensive and robust supernova compilations currently available in the literature, while Union2.1 is selected because it is characterized by larger statistical uncertainties in the luminosity distance measurements, allowing us to assess uncertainties in the value of $H_0$ that propagate into the VSL parameters. This motivates our choice of Pantheon$+$ and Union2.1 as two datasets that maximize the allowed variation in $H_0$ while remaining statistically consistent.

It is important to emphasize that, despite this difference in best-fit values, all late-time determinations of $H_0$ from supernova data remain mutually consistent at the $1\sigma$ level. The observed spread should therefore be interpreted as residual statistical uncertainty rather than as evidence for a genuine tension among late-time datasets. In this sense, late-Universe probes based on supernovae are fully compatible  with each other. However, since the central modification in our model affects only the luminosity distance, it is precisely this residual uncertainty in $d_L$ that can propagate into the parameters describing the variation of the speed of light.

\paragraph*{Discrepancy by BAO:}
The origin of the discrepancy that emerges once BAO data are included can be understood in terms of the role played by absolute versus relative distance measurements. The standard determination of the Hubble constant $H_0$ relies on the distance ladder, in which distances in the nearby Universe are used to calibrate secondary indicators that are then applied to galaxies in the Hubble flow. In this approach, $H_0$ is inferred from the relation $H_0 = v/d$, where peculiar velocities are subdominant at sufficiently large distances.

Baryon Acoustic Oscillations provide an alternative and complementary strategy. 
Since the BAO scale is set by well-understood early-Universe physics, the sound 
horizon $r_s$ can be computed in absolute units. As a consequence, the combination 
of BAO with Type Ia supernovae enables an ``inverse distance ladder,'' in which BAO 
fix the absolute scale while supernovae provide a relative distance--redshift 
relation extending to low redshifts. In this framework, the Hubble constant is 
effectively determined by the slope of the late-time distance--redshift relation. See Ref.~\cite{BOSS:2014hhw} for a more detailed discussion.

The impact of this procedure depends sensitively on the statistical precision of 
the supernova sample employed. The Pantheon$+$ compilation provides a highly 
precise relative distance scale, allowing the BAO absolute ruler to be smoothly 
and efficiently propagated to low redshifts. As a result, the inferred value of 
$H_0$ is largely anchored by the supernova data, with BAO acting primarily to set 
the overall normalization.

In contrast, the Union2.1 dataset is characterized by larger statistical 
uncertainties in the luminosity distance measurements. In this case, the weighted 
combination of datasets assigns a more significant role to the BAO observations, 
which typically carry smaller relative uncertainties. Consequently, the BAO scale 
exerts a stronger pull on the inferred value of $H_0$, shifting its central value 
toward $H_0 \simeq 69.7~\mathrm{km\,s^{-1}\,Mpc^{-1}}$ when Union2.1 is combined with 
BAO data.

\paragraph*{Uncertainty in $H_{0}$ and $c(z)$:}
There might be a concern regarding the interpretation of our results 
and the apparent difference between the conclusions obtained using Pantheon$+$  and Union2.1. The two supernova compilations explore different realizations of the same 
late-Universe uncertainty in the determination of the Hubble constant. This behavior is well understood in inverse distance ladder analyses and reflects differences in statistical precision rather than an inconsistency between the datasets.

In our Brans--Dicke-like VSL framework, this uncertainty in $H_0$ is unavoidably 
transmitted to the parameterization of $c(z)$. Since the luminosity distance is 
the only cosmological observable that departs from its $\Lambda$CDM expression, 
any shift in the BAO-anchored value of $H_0$ propagates directly into the inferred 
variation of the speed of light. Consequently, a higher value of $H_0$, as favored 
by Pantheon$+$ \& BAO, translates into a varying $c(z)$, whereas a lower 
value of $H_0$, as favored by Union2.1 \& BAO, naturally leads to consistency with a 
constant speed of light. To visualize this behavior, it suffices to substitute Eq.~(40) into Eq.~(45), from which it becomes evident how the dependence on $H_0$ directly impacts the parameters governing $c(z)$.

The main result is that this shift in the inferred value of $H_0$ is directly transmitted to the parameterization of $c(z)$, revealing a correlation that emerges between $H_0$ and VSL. This correlation implies that the current uncertainty in $H_0$ impacts not only $\Lambda$CDM inferences but also beyond-$\Lambda$CDM sectors in which $G$ and $c$ co-vary.

\section{Final remarks}\label{sec:Final-remarks}

In this paper we have constrained a Brans-Dicke-like model for co-varying $G$ and $c$ using observational data. We adopted three different parameterizations for the $c=c\left(z\right)$, namely: power-law (for its simplicity); Gupta's ansatz (due to phenomenological sucess); continuous (because of no-need for a cut-off). The datasets utilized were (a combination of) Pantheon$+$ (related to SN Ia observations), Union2.1 (SN Ia), DESI (BAO data), and $\theta_*$ (CMB data from Planck). 

In the BD-like VSL framework, several cosmological distances remain unchanged with respect to the standard $\Lambda$CDM model. The proper distance, the angular diameter distance, and sound-horizon-related distances follow the same expressions as in standard cosmology. The only modification arises in the luminosity distance $d_{L}$, which has an additional
multiplicative factor $\left(c/c_{0}\right)$, as shown in Section \ref{subsec:Luminosity-distance}. Since supernova type Ia data are directly tied to $d_{L}$, it is precisely through this dataset that deviations from the standard cosmology are expected to manifest.

The Pantheon+ dataset points to a strong preference for a varying speed of light, with evidence at the $3\sigma$ level across all three parameterizations of $c(z)$ that we investigated. In contrast, the Union2.1 compilation favors instead the standard constant-$c$ case. This divergence arises entirely from the change in the SN Ia dataset employed, which is fully consistent with the theoretical structure of our model: since the luminosity distance is the only quantity in which our scenario departs from $\Lambda$CDM, it is precisely the supernova observations that determine whether or not deviations are detected.

As already established in the literature, combining Pantheon$+$ with BAO data or Union2.1 with BAO leads to significantly different estimates of $H_{0}$. In our analysis, we have shown that this shift in the inferred value of $H_{0}$ is directly transmitted to the parameterization of $c(z)$, revealing a correlation that emerges between $H_{0}$ and VSL. This correlation implies that the present $H_{0}$ tension impacts not only $\Lambda$CDM inferences but also beyond-$\Lambda$CDM sectors where $G$ and $c$ co-vary. In our BD-like VSL runs, a high-$H_{0}$ value (e.g., $H_{0}\simeq 73\,\mathrm{km\,s^{-1}\,Mpc^{-1}}$) drives the posteriors toward a non-zero $c(z)$ with significance $\gtrsim 3\sigma$ across all three parameterizations, whereas a lower value (e.g., $H_{0}\simeq 70\,\mathrm{km\,s^{-1}\,Mpc^{-1}}$) renders the constant-$c$ limit fully consistent with the data. This behavior reflects the fact that SNe~Ia primarily constrains the combination $d_{L}\propto (c/c_{0})\,H_{0}^{-1}$ at $z\lesssim 1$: decreasing $H_{0}$ can be partially compensated by $c(z_{\rm SN})<c_{0}$, and vice versa. Consequently, quantitative statements about BD-like VSL are currently \emph{$H_{0}$-limited}.

From a further perspective, two complementary avenues can refine and stress-test these results: (i) enlarge the data vector with late-time, SN-independent distance anchors---strong-lensing time delays, standard sirens, water megamasers and cosmic-chronometer $H(z)$---together with homogeneous SN calibration, to break the $(c/c_{0})H_{0}^{-1}$ degeneracy; and (ii) widen the theory space by analyzing other classes of VSL different from BD-like realization. Whether the correlation between $H_{0}$ and VSL persists across these tests will indicate whether it is a generic phenomenological degeneracy or a model-dependent structure that forthcoming data can discriminate.

\section*{Acknowledgements}

The authors thank Prof. Rajendra P. Gupta for useful discussions. JBS acknowledges CNPq-Brazil for its support. RRC and LGM thank CNPq-Brazil for partial financial support---Grants: 309063/2023-0 (RRC) and 307901/2022-0 (LGM). RRC is grateful to FAPEMIG-Brazil (Grants APQ-00544-23 and APQ-05218-23) for financial support. PJP acknowledges LAB-CCAM at ITA and CNPq (Grant 401565/2023-8). The authors thank the referee for the valuable suggestions that helped us improve the text.

\section*{Data Availability}

As mentioned before, this papers makes use of observational data from the Pantheon+ \citep[]{Pantheon+}, Union2.1 \citep[]{Union2.1}, and DESI \citep[]{DESI2025EDE} collaborations. All datasets are publicly available and were provided by the respective collaborations. Union2.1 data are available on the project website: https://supernova.lbl.gov/union. The Pantheon+ data can be accessed through the official GitHub repository: https://github.com/PantheonPlusSH0ES/DataRelease. DESI data were obtained directly from the original publication by the collaboration.

\appendix

\section{The redshift in varying-$c$ frameworks }\label{app:Redshift}

Consider Eq. (\ref{eq:r(t)}). The distance $r$ travelled by a wave
crest from emission at $t_{e}$ until detection at $t_{0}$ is: 
\[
r={\displaystyle \int_{t_{e}}^{t_{0}}}\frac{c\left(t\right)dt}{a\left(t\right)}.
\]
It is the same as the distance $r$ travelled by the next wave crest
from emission at $t_{e}+\lambda_{e}/c_{e}$ until detection at $t_{0}+\lambda_{0}/c_{0}$,
\[
r={\displaystyle \int_{t_{e}+\lambda_{e}/c_{e}}^{t_{0}+\lambda_{0}/c_{0}}}\frac{c\left(t\right)dt}{a\left(t\right)},
\]
because we assume that neither the universe nor the speed of light
has had time to change significantly in such a reduced time as that
between two subsequent wave crests. The symbol $\lambda$ stands for
the radiation's wavelength. For future reference, $\nu$ is the frequency
of the photon associated to the radiation.

Equating the previous results: 
\[
{\displaystyle \int_{t_{e}}^{t_{0}}}\frac{c\left(t\right)dt}{a\left(t\right)}={\displaystyle \int_{t_{e}+\lambda_{e}/c_{e}}^{t_{0}+\lambda_{0}/c_{0}}}\frac{c\left(t\right)dt}{a\left(t\right)}
\]
Subtracting 
\[
{\displaystyle \int_{t_{e}+\lambda_{e}/c_{e}}^{t_{0}}}\frac{c\left(t\right)dt}{a\left(t\right)}
\]
from the equation above leads to: 
\[
{\displaystyle \int_{t_{e}}^{t_{e}+\lambda_{e}/c_{e}}}\frac{c\left(t\right)dt}{a\left(t\right)}={\displaystyle \int_{t_{0}}^{t_{0}+\lambda_{0}/c_{0}}}\frac{c\left(t\right)dt}{a\left(t\right)}.
\]
Assuming $a\simeq a_{0}$ and $c=c_{0}$ between the time of detection
of one wave crest and the next (and the same argument for the emission
of subsequent wave crests), 
\[
{\displaystyle \int_{t_{e}}^{t_{e}+\lambda_{e}/c_{e}}}\frac{c_{e}dt}{a_{e}}={\displaystyle \int_{t_{0}}^{t_{0}+\lambda_{0}/c_{0}}}\frac{c_{0}dt}{a_{0}}\Rightarrow\frac{c_{e}}{a_{e}}{\displaystyle \int_{t_{e}}^{t_{e}+\lambda_{e}/c_{e}}}dt=\frac{c_{0}}{a_{0}}{\displaystyle \int_{t_{0}}^{t_{0}+\lambda_{0}/c_{0}}}dt,
\]
which leads to 
\begin{equation}
\frac{\lambda_{e}}{a_{e}}=\frac{\lambda_{0}}{a_{0}}.\label{eq:l_0(l_e)}
\end{equation}

The redshift $z$ is defined as the fractional change in the wavelength
of the radiation: 
\begin{equation}
z\equiv\frac{\lambda_{0}-\lambda_{e}}{\lambda_{e}}.\label{eq:z(l_0,l_e)}
\end{equation}
It is then expressed in terms of the scale factor via (\ref{eq:l_0(l_e)}):
\begin{equation}
z\equiv\frac{\lambda_{0}-\frac{a_{e}}{a_{0}}\lambda_{0}}{\frac{a_{e}}{a_{0}}\lambda_{0}}=\frac{\left(1-\frac{a_{e}}{a_{0}}\right)}{\frac{a_{e}}{a_{0}}}=\frac{a_{0}}{a_{e}}-1.\label{eq:z(a_e)}
\end{equation}
Henceforth $a_{e}=a$. Therefore, the expression for the redshift
in terms of the scale factor, 
\begin{equation}
\left(1+z\right)=\frac{a_{0}}{a}\label{eq:z(a)-app}
\end{equation}
is still valid in the context of varying-$c$ scenarios. Eq. (\ref{eq:z(a)-app})
is precisely Eq. (\ref{eq:z(a)}) with $a_{0}\equiv1$.

All the above consider the following two assumptions: (i) the value of $c_{1}$ at the time of emission of the light front in the cosmic void is equal to the value of $c_{SN}$ in the host galaxy of the source; (ii) the value of $c_{2}$ at current time in the cosmic void is equals to $c_{MW}$ inside the Milky Way (hosting the observers). However, it is a known fact that matter is stabilized inside galaxies which resist cosmic expansion. This matter could, in principle, induce refraction, altering the wavelength $\lambda$ in the passage from a medium to the next (see Figure 1 of \cite{Nguyen2026}). This would be consistent with
\[
\frac{\lambda_{1}}{\lambda_{SN}}=\frac{c_{1}}{c_{SN}}\qquad\rm{and}\qquad\frac{\lambda_{MW}}{\lambda_{2}}=\frac{c_{MW}}{c_{2}}.
\]
For this reason, the redshift formula would read
\[
1+z=\frac{\lambda_{MW}}{\lambda_{SN}}=\frac{\lambda_{MW}}{\lambda_{2}}\frac{\lambda_{2}}{\lambda_{1}}\frac{\lambda_{1}}{\lambda_{SN}}=\frac{c_{MW}}{c_{2}}\frac{a_{0}}{a}\frac{c_{1}}{c_{SN}}=\frac{a_{0}}{a}\frac{c_{1}}{c_{2}}\frac{c_{MW}}{c_{SN}}.
\]
One could use the modified Lemaitre formula above to extend the treatment developed in this paper. This is outside the scope of the present work and could be done elsewhere in the future.\footnote{The authors thank the referee for the remark on the possibility of a modified Lemaitre formula.}

%%%%%%%%%%%%%%%%%%%%%
%    Bibliography   %
%%%%%%%%%%%%%%%%%%%%%

\end{document}